\documentclass[11pt,a4paper]{article}
\pdfoutput=1

\usepackage{jheppub}

\usepackage{amsmath}
\usepackage{bbm}
\usepackage{float}
\usepackage{graphicx}	
\usepackage{hyperref}
\usepackage{mathtools}
\usepackage{cancel}

\usepackage[normalem]{ulem}

\usepackage{multirow}
\usepackage{rotating}
\usepackage{subcaption}

\def\lsim{\raise0.3ex\hbox{$\;<$\kern-0.75em\raise-1.1ex
\hbox{$\sim\;$}}}
\def\gsim{\raise0.3ex\hbox{$\;>$\kern-0.75em\raise-1.1ex
\hbox{$\sim\;$}}}

\def\thetitle{ 
Neutrino amplitude decomposition, $S$ matrix rephasing invariance, and reparametrization symmetry \\
 \vspace{- 6mm}
}
\title{\thetitle}
\hypersetup{pdftitle={\thetitle}}
\author{Hisakazu Minakata}
\affiliation{
Center for Neutrino Physics, Department of Physics, Virginia Tech, Blacksburg, Virginia 24061, USA \\
}

\emailAdd{hisakazu.minakata@gmail.com}
\date{\today}

\abstract{
The $S$ matrix rephasing invariance is one of the fundamental principles of quantum mechanics that originates in its probabilistic interpretation. For a given $S$ matrix which describes neutrino oscillation, one can define the two different rephased amplitudes $S_{\alpha \beta}^{ \text{Reph-1} } \equiv e^{ i (\lambda_{1} / 2E) x} S_{\alpha \beta}$ and $S_{\alpha \beta}^{ \text{Reph-2} } \equiv e^{ i (\lambda_{2} / 2E) x} S_{\alpha \beta}$, which are physically equivalent to each other, where $\lambda_{k} / 2E$ denotes the energy eigenvalue of the $k$-th mass eigenstate. We point out that the transformation of the reparametrization (Rep) symmetry obtained with ``Symmetry Finder'' maps $S_{\alpha \beta}^{ \text{Reph-1} }$ to $S_{\alpha \beta}^{ \text{Reph-2} }$, and vice versa, providing a local and manifest realization of the $S$ matrix rephasing invariance by the Rep symmetry of the 1-2 state exchange type. It is strongly indicative of quantum mechanical nature of the Rep symmetry. The rephasing and Rep symmetry relation, though its all-order treatment remains incomplete, is shown to imply absence of the pure 1-3 exchange symmetry in Denton~{\it et al.}~perturbation theory. It then triggers a study of convergence of perturbation series. 
}

\begin{document} % JHEP 

\maketitle

\section{Introduction}
\label{sec:introduction} 

Theory of neutrino oscillation heavily relies on quantum mechanics~\cite{Dirac:1958}. Without the concept of superposition of states, even the first page of a textbook describing this phenomenon cannot be written. The theoretical framework of quantum mechanics, which predicts the wave function of the states, or the transition $S$ matrix, meets the real world through the probabilistic interpretation. One of the consequences of the probabilistic nature of the theory is that the oscillation $S$ matrix is invariant under the global or local rephasing transformations, which we refer as the $S$ matrix rephasing invariance in this paper. 

The rephasing factor of the $S$ matrix which leaves the physical observables invariant may have a non-local character, $S \rightarrow S e^{i \int_{0}^{x} dx^{\prime} \mathcal{A} (x^{\prime}) }$, which depends on global property of the variable $\mathcal{A} (x)$. This type of rephasing is used, for example, to remove the Wolfenstein matter potential~\cite{Wolfenstein:1977ue} due to neutral current interactions from the evolution equation in the discussion of the MSW effect~\cite{Wolfenstein:1977ue,Mikheyev:1985zog}, or the matter-enhanced oscillations~\cite{Barger:1980tf}, for neutrinos traversing in a uniform- or nonuniform-density matter. 

In the other case, the $S$ matrix rephasing factor can take a local form but depends upon the dynamical variables, i.e., the eigenvalues of the quantum operators~\cite{Dirac:1958}. In section~\ref{sec:amp-decomposition}, we discuss physical equivalence between the two rephased oscillation amplitudes $S_{\alpha \beta}^{ \text{Reph-1} } \equiv e^{ i ( \lambda_{1} / 2E) x} S_{\alpha \beta}$ and $S_{\alpha \beta}^{ \text{Reph-2} } \equiv e^{ i ( \lambda_{2} / 2E) x} S_{\alpha \beta}$ for a given $S$ matrix elements $S_{\alpha \beta}$, where $\lambda_{k} / 2E$ denote the energy eigenvalues of the $k$-th mass eigenstates in matter with $E$ being the neutrino energy. We will let the readers know how the two differently rephased amplitudes came to our attention. As it stands the realization of the $S$ matrix rephasing is quite broad in its form. 
We use the Greek indices $\alpha, \beta$ etc. for the neutrino flavor indices and the Latin ones $i, j, k$ for the mass eigenstate indices throughout this paper. 

In this paper, we want to provide a new way of realizing the $S$ matrix rephasing invariance in a manner which we phrase ``locally and manifestly''. If the symmetry transformation exists for the dynamical variables collectively denoted as $\Phi$ which map $S^{ \text{Reph-1} }$ to $S^{ \text{Reph-2} }$, 
\begin{eqnarray} 
S_{\alpha \beta}^{ \text{Reph-1} } \vert_{ \Phi \rightarrow \Phi^{\prime} } \rightarrow S_{\alpha \beta}^{ \text{Reph-2} }, 
\label{Phi-transf} 
\end{eqnarray}
or vice versa, we say that the $S$ matrix rephasing invariance is represented locally and manifestly. The mapping is ``local'' because the two rephased $S$ matrices which differ by position-dependent factors are related to each other by the $\Phi$ symmetry transformations at each spatial point. It is ``manifest'' because the physical equivalence between the two rephased $S$ matrices becomes obvious without computing the probability. 
What is remarkable is that the reparametrization symmetry in neutrino oscillation in matter discussed in refs.~\cite{Minakata:2021dqh,Minakata:2021goi,Minakata:2022zua,Minakata:2022yvs,Minakata:2022pyr} plays the role of the $\Phi$ symmetry, as in eq.~\eqref{Phi-transf}. For our frequent usage, we denote the reparametrization symmetry discussed in these references as the ``Rep symmetry'' throughout this paper. 

A few words for the Rep symmetry in neutrino oscillation in matter: 
Recently a systematic search for the Rep symmetry is carried out by using the method called ``Symmetry Finder'' (SF)~\cite{Minakata:2021dqh,Minakata:2021goi,Minakata:2022zua,Minakata:2022yvs,Minakata:2022pyr}. It has been started from the heuristic observation in vacuum~\cite{Parke:2018shx} that the flavor basis state (i.e., wave function) $\nu$ allows the two different expressions by the mass eigenstate $\bar{\nu}$ in vacuum and in matter as~\cite{Minakata:2021dqh}  
\begin{eqnarray} 
&&
\nu = U (\theta_{23}, \theta_{13}, \theta_{12}, \delta) \bar{\nu} 
= U (\theta_{23}^{\prime}, \theta_{13}^{\prime}, \theta_{12}^{\prime}, \delta^{\prime}) \bar{\nu}^{\prime}, 
\label{SF-eq-general}
\end{eqnarray}
where $U \equiv U_{\text{\tiny MNS}}$ denotes the standard three-flavor neutrino mixing matrix~\cite{Maki:1962mu}. In matter the $U$ matrix must be replaced by the $V$ matrix~\cite{Minakata:1998bf} with the matter-affected mixing angles and the states defined with the matter effect. For explicit examples of the $V$ matrix computation of the oscillation probability, see refs.~\cite{Minakata:2015gra,Denton:2016wmg,Minakata:2022zua}. Since eq.~\eqref{SF-eq-general}, or $U \rightarrow V$ version in matter, represents the unique flavor state by the energy eigenstates with two different sets of the physical parameters, it implies a symmetry. This is nothing but the key statement of SF~\cite{Minakata:2021dqh,Minakata:2021goi,Minakata:2022zua,Minakata:2022yvs,Minakata:2022pyr,Parke:2018shx}. 

Up to now we have only a handful success examples of our symmetry hunting. Around the solar resonance we have uncovered the twin~\cite{Minakata:2022pyr} eight 1-2 state exchange Rep symmetries~\cite{Minakata:2021dqh,Minakata:2022yvs} in the solar-resonance perturbation (SRP) theory~\cite{Martinez-Soler:2019nhb} and Denton {\it et al.} (DMP) perturbation theory~\cite{Denton:2016wmg}. Around the atmospheric resonance we have the so-far-unique sixteen 1-3 state exchange symmetry~\cite{Minakata:2021goi} in the helio-perturbation theory~\cite{Minakata:2015gra}, ``unique'' because we suffer from the problem of missing 1-3 symmetry in DMP as discussed in section~\ref{sec:rephasing-wider}.  
Quite recently we have made a progress in digging out the full eight 1-2 exchange symmetries in vacuum~\cite{Minakata:2022zua}, completing the first treatment done in ref.~\cite{Parke:2018shx}. It implies the similar eight 1-2 exchange symmetries,  Symmetry X-ZS, in the Zaglauer-Schwarzer (ZS) construction~\cite{Zaglauer:1988gz} in matter with the vacuum variables fully replaced by the matter-affected ones~\cite{Minakata:2022zua}.

In this paper, for clarity, we restrict our discussion to the neutrino-mass-embedded Standard Model ($\nu$SM). Throughout this paper we discuss intimate relationship between the $S$ matrix rephasing invariance and the Rep symmetry. It will illuminate the quantum-mechanics-rooted nature of the Rep symmetry, and clarifies in which way this property manifests itself. We believe that it implies a clear answer to the question repeatedly raised in our previous references: What is the raison d'~\^etre of the Rep symmetry? It exists to provide a local and manifest realization of the $S$ matrix rephasing invariance. This observation is in perfect harmony with the quantum nature of the Rep symmetry emphasized in ref.~\cite{Minakata:2022zua}. 
Finally, we show that the newly gained viewpoint ``Rep symmetry from rephasing invariance'' reveals otherwise concealed nature of the Rep symmetry, including absence of the pure 1-3 state exchange symmetry in DMP. 

\section{Neutrino amplitude decomposition}
\label{sec:amp-decomposition} 

We have encountered the local realization of the $S$ matrix rephasing invariance, eq.~\eqref{Phi-transf}, in a seemingly completely unrelated context of looking for the way of how to decompose the oscillation $S$ matrix into the atmospheric- and the solar-scale oscillation waves consistently~\cite{Minakata:2020oxb}. It is to prepare the theoretical framework for observing the interference effects between the solar- and atmospheric-scale amplitudes by using the experimental data. In fact, the amplitude decomposition has been formulated first in vacuum and applied~\cite{Huber:2019frh} to a future reactor neutrino data taken by a JUNO~\cite{JUNO:2022mxj}-like detector, an ideal setting for the interference analysis with a high-performance detector sensitive to the both solar- and the atmospheric-scale oscillations. 
Then, we tried to incorporate the accelerator long-baseline experiments and the atmospheric neutrino observation into the amplitude decomposition formalism~\cite{Minakata:2020ijz,Minakata:2020oxb}. We note that we are heading the era in which we will be able to observe such an interference effect by analyzing the data taken by the ongoing and upcoming neutrino experiments~\cite{JUNO:2022mxj,T2K:2023mcm,NOvA:2021nfi,Hyper-Kamiokande:2018ofw,DUNE:2020jqi,IceCubeCollaboration:2023wtb,KM3NeT:2021ozk}. 

To make the discussion on the rephasing invariance a concrete one, we take the particular framework, the DMP perturbation theory~\cite{Denton:2016wmg}, for description of neutrino oscillation in matter. In this section, we recollect the relevant part of discussion of the amplitude decomposition~\cite{Minakata:2020oxb}. For simplicity and clarity of our discussion, we take the uniform matter density approximation in this paper. The flavor basis $S$ matrix elements in matter can be written as 
\begin{eqnarray} 
S_{\alpha \beta} &=& 
V_{\alpha 1} V^{*}_{\beta 1} e^{ - i \frac{ \lambda_{1} }{2 E} x}  
+ V_{\alpha 2} V^{*}_{\beta 2} e^{ - i \frac{ \lambda_{2} }{2 E} x}   
+ V_{\alpha 3} V^{*}_{\beta 3} e^{ - i \frac{ \lambda_{3} }{2 E} x}, 
\hspace{3mm} 
\label{S-matrix-matter}
\end{eqnarray}
where $\lambda_{i} / 2E$ ($i=1,2,3$) denote the energy eigenvalues in matter. In DMP, the $V$ matrix elements~\cite{Minakata:1998bf} are computed to first a few orders of the DMP perturbation theory~\cite{Denton:2016wmg}. Using unitarity of the $V$ matrix, it is shown that the decomposition of the $S$ matrix into the ones with the two frequencies, $S_{\alpha \beta} = \delta_{\alpha \beta} + S_{\alpha \beta}^{ \text{atm} } + S_{\alpha \beta}^{ \text{sol} }$, can be carried out~\cite{Minakata:2020ijz,Minakata:2020oxb}. 

\subsection{Two rephased amplitudes $S_{\alpha \beta}^{ \text{Reph-1} }$ and $S_{\alpha \beta}^{ \text{Reph-2} }$ }
\label{sec:2-amplitudes} 

For simplicity we restrict ourselves to the appearance channel $\alpha \neq \beta$, for which unitarity becomes $\sum_{i} V_{\alpha i} V^{*}_{\beta i} = 0$. One can decompose the $S$ matrix in the two different ways as 
\begin{eqnarray} 
&& 
S_{\alpha \beta} = 
e^{ - i \frac{ \lambda_{1} }{2E} x} 
\left[ 
V_{\alpha 2} V^{*}_{\beta 2} 
\left\{ e^{ - i \frac{ ( \lambda_{2} - \lambda_{1} ) }{2E} x} - 1 \right\} 
+ V_{\alpha 3} V^{*}_{\beta 3} 
\left\{ e^{ - i \frac{ ( \lambda_{3} - \lambda_{1} ) }{2E} x} - 1 \right\} 
\right], 
\nonumber \\
&& 
S_{\alpha \beta} = 
e^{ - i \frac{ \lambda_{2} }{2E} x} 
\left[ 
V_{\alpha 1} V^{*}_{\beta 1} 
\left\{ e^{ i \frac{ ( \lambda_{2} - \lambda_{1} ) }{2E} x} - 1 \right\} 
+ V_{\alpha 3} V^{*}_{\beta 3} 
\left\{ e^{ - i \frac{ ( \lambda_{3} - \lambda_{2} ) }{2E} x} - 1 \right\} 
\right], 
\label{amp-decompose-2ways}
\end{eqnarray}
where the first term in each parenthesis involving the $e^{ \pm i \frac{ ( \lambda_{2} - \lambda_{1} ) }{2E} x}$ factor represents the solar-scale oscillation, and the second one with $e^{ - i \frac{ ( \lambda_{3} - \lambda_{k} ) }{2E} x}$ (k=1,2) the atmospheric-scale one. Thus, we have obtained the two different rephased $S$ matrices which are decomposed into the solar and atmospheric waves as in eq.~\eqref{amp-decompose-2ways}: 
\begin{eqnarray} 
&&
S_{\alpha \beta}^{ \text{Reph-1} } \equiv 
e^{ i \frac{ \lambda_{1} }{2E} x} S_{\alpha \beta}, 
\hspace{10mm} 
S_{\alpha \beta}^{ \text{Reph-2} } \equiv 
e^{ i \frac{ \lambda_{2} }{2E} x} S_{\alpha \beta}. 
\label{S-Reph-12-def} 
\end{eqnarray} 
By being different only by the overall phase, of course, these two rephased amplitudes lead to the identical oscillation probability, the observable. 

\subsection{DMP: Basic notations}
\label{sec:notations} 

To write down the explicit form of $S_{\alpha \beta}^{ \text{Reph-1} }$ and $S_{\alpha \beta}^{ \text{Reph-2} }$, we need a minimal exposition of the notations. In DMP~\cite{Denton:2016wmg}  the mixing angles $\theta_{12}$ and $\theta_{13}$ in vacuum are elevated to $\psi$ and $\phi$ due to the matter effect: They are the matter-affected $\theta_{12}$ and $\theta_{13}$, respectively, and $s_{12}$ and $s_{\psi}$ etc. are shorthand notations for $\sin \theta_{12}$ and $\sin \psi$ etc.. A minimal recollection of the DMP perturbation theory will be given in section~\ref{sec:recollection-DMP}. The Hamiltonian eq.~\eqref{barH-0th-1st} defines the perturbation theory with $\epsilon$, the unique expansion parameter which is defined as 
\begin{eqnarray} 
&&
\epsilon \equiv \frac{ \Delta m^2_{21} }{ \Delta m^2_{ \text{ren} } }, 
\hspace{10mm}
\Delta m^2_{ \text{ren} } \equiv \Delta m^2_{31} - s^2_{12} \Delta m^2_{21},
\label{epsilon-Dm2-ren-def}
\end{eqnarray}
where $\Delta m^2_{ \text{ren} }$ is the ``renormalized'' atmospheric $\Delta m^2$ used in ref.~\cite{Minakata:2015gra}.\footnote{
%%%%%%%%%%%%%% footnote %%%%%%%%%%%%%%
A debate~\cite{Minakata:2015gra} on preference of this notation $\Delta m^2_{ \text{ren} }$ or the other option $\Delta m^2_{ \text{ee} }$~\cite{Nunokawa:2005nx} still continues. }
To keep the expressions of the $S$ matrices as simple as possible, we introduce the new notations for the zeroth-order eigenvalues in the Hamiltonian~\eqref{barH-0th-1st} and the $2E$-scaled $\Delta m^2_{ \text{ren} }$:
\begin{eqnarray} 
&&
h_{i} \equiv \frac{ \lambda_{i} }{2E}, 
\hspace{10mm} 
\Delta_{ \text{ren} } \equiv 
\frac{ \Delta m^2_{ \text{ren} } }{2E}, 
\label{hi-def} 
\end{eqnarray}
which is to be used only in the expressions of the $S$ matrices. 
For the mixing matrix, we use the SOL convention $U$ matrix~\cite{Parke:2018shx,Martinez-Soler:2018lcy}. It is defined by the phase redefinition of $U_{\text{\tiny PDG}}$ of the Particle Data Group (PDG)~\cite{ParticleDataGroup:2022pth} convention: 
\begin{eqnarray} 
U_{\text{\tiny SOL}} 
&=&
\left[
\begin{array}{ccc}
1 & 0 &  0  \\
0 & e^{ - i \delta} & 0 \\
0 & 0 & e^{ - i \delta} \\ 
\end{array}
\right] 
U_{\text{\tiny PDG}} 
\left[
\begin{array}{ccc}
1 & 0 &  0  \\
0 & e^{ i \delta} & 0 \\
0 & 0 & e^{ i \delta} \\
\end{array}
\right] 
=
\left[
\begin{array}{ccc}
1 & 0 &  0  \\
0 & c_{23} & s_{23} \\
0 & - s_{23} & c_{23} \\
\end{array}
\right] 
\left[
\begin{array}{ccc}
c_{13}  & 0 & s_{13} \\
0 & 1 & 0 \\
- s_{13} & 0 & c_{13} \\
\end{array}
\right] 
\left[
\begin{array}{ccc}
c_{12} & s_{12} e^{ i \delta}  &  0  \\
- s_{12} e^{- i \delta} & c_{12} & 0 \\
0 & 0 & 1 \\
\end{array}
\right] 
\nonumber \\
&\equiv& 
U_{23} (\theta_{23}) U_{13} (\theta_{13}) U_{12} (\theta_{12}, \delta). 
\label{U-SOL-def} 
\end{eqnarray}
The second line in eq.~\eqref{U-SOL-def} defines, in order, the 2-3, 1-3, and 1-2 rotation matrices. In $U_{\text{\tiny SOL}}$ the CP phase factor $e^{ \pm i \delta}$ is attached to the solar angle $\sin \theta_{12}$, hence the convention is termed as ``SOL''. By being related by the phase redefinition, of course, the oscillation probability computed with the SOL convention $U_{\text{\tiny SOL}}$ matrix is identical with that obtained with $U_{\text{\tiny PDG}}$. 

\subsection{$S$ matrix and rephased amplitudes in the $\nu_{\mu} \rightarrow \nu_{e}$ channel} 
\label{sec:S-rephased-emu}

To bring a little more intuition to our discussion, we present the expressions of the $S$ matrix elements in the $\nu_{\mu} \rightarrow \nu_{e}$ channel. We do not plan to go into detail of the computation as it was carried out in ref.~\cite{Denton:2016wmg}. If the readers prefer, see also ref.~\cite{Minakata:2020oxb} for the less abstract expressions of the $S$ matrices. One can compute the $S$ matrix element $S_{e \mu}$ in the $\nu_{\mu} \rightarrow \nu_{e}$ channel to first order in the DMP expansion, and the result can be written, using the notation $s_{ (\phi - \theta_{13}) } \equiv \sin (\phi - \theta_{13})$, as 
\begin{eqnarray} 
S_{e \mu} 
%\nonumber \\
&=& 
c_{23} c_{\phi} e^{ i \delta} 
c_\psi s_\psi \left( e^{ - i h_{2} x }  - e^{ - i h_{1} x } \right) 
+ s_{23} c_{\phi} s_{\phi} 
\left[ e^{ - i h_{3} x } - \left( c^2_\psi e^{ - i h_{1} x } + s^2_\psi e^{ - i h_{2} x } \right) 
\right] 
\nonumber \\
&+& 
\epsilon c_{12} s_{12} s_{ (\phi - \theta_{13}) } 
\left( 
c_{23} s_{\phi} e^{ i \delta} c^2_\psi + s_{23} \cos 2\phi c_{\psi} s_\psi 
\right) 
\frac{ \Delta_{ \text{ren} } }{ h_{3} - h_{2} } 
\left( e^{ - i h_{3} x } - e^{ - i h_{2} x } \right) 
\nonumber \\
&+& 
\epsilon c_{12} s_{12} s_{ (\phi - \theta_{13}) } 
\left( 
c_{23} s_{\phi} e^{ i \delta} s^2_\psi - s_{23} \cos 2\phi c_{\psi} s_\psi 
\right) 
\frac{ \Delta_{ \text{ren} } }{ h_{3} - h_{1} } 
\left( e^{ - i h_{3} x } - e^{ - i h_{1} x } \right). 
%
%\nonumber \\
\label{S-emu}
\end{eqnarray}
Then, using the definitions in eq.~\eqref{S-Reph-12-def}, the rephased amplitudes $S_{e \mu}^{ \text{Reph-1} }$ and $S_{e \mu}^{ \text{Reph-2} }$ can readily be obtained as 
\begin{eqnarray} 
\hspace{-8mm}
S_{e \mu}^{ \text{Reph-1} } 
&=&
s_{23} c_{\phi} s_{\phi} \left\{ e^{ - i ( h_{3} - h_{1} ) x } - 1 \right\}  
+ c_{\phi} \left( c_{23} c_\psi s_\psi e^{ i \delta} - s_{23} s_{\phi} s^2_\psi \right) 
\left\{ e^{ - i ( h_{2} - h_{1} ) x } - 1 \right\} 
\nonumber \\
&-& 
\epsilon c_{12} s_{12} s_{ (\phi - \theta_{13}) } 
\left( 
c_{23} s_{\phi} c^2_\psi e^{ i \delta} + s_{23} \cos 2\phi c_{\psi} s_\psi 
\right) 
\frac{ \Delta_{ \text{ren} } }{ h_{3} - h_{2} } 
\left\{ e^{ - i ( h_{2} - h_{1} ) x } - 1 \right\} 
\nonumber \\
&+& 
\epsilon c_{12} s_{12} s_{ (\phi - \theta_{13}) } 
\left( 
c_{23} s_{\phi} c^2_\psi e^{ i \delta} + s_{23} \cos 2\phi c_{\psi} s_\psi 
\right) 
\frac{ \Delta_{ \text{ren} } }{ h_{3} - h_{2} } 
\left\{ e^{ - i ( h_{3} - h_{1} ) x } -1 \right\} 
\nonumber \\
&+& 
\epsilon c_{12} s_{12} s_{ (\phi - \theta_{13}) } 
\left( 
c_{23} s_{\phi} s^2_\psi e^{ i \delta} - s_{23} \cos 2\phi c_{\psi} s_\psi 
\right) 
\frac{ \Delta_{ \text{ren} } }{ h_{3} - h_{1} } 
\left\{ e^{ - i ( h_{3} - h_{1} ) x } - 1 \right\}. 
\label{S-emu-Reph1}
\end{eqnarray}
\begin{eqnarray} 
\hspace{-8mm}
S_{e \mu}^{ \text{Reph-2} } 
&=&
s_{23} c_{\phi} s_{\phi} \left\{ e^{ - i ( h_{3} - h_{2} ) x } - 1 \right\} 
- c_{\phi} \left( 
c_{23} c_\psi s_\psi e^{ i \delta} 
+ s_{23} s_{\phi} c^2_\psi \right) 
\left\{ e^{ i ( h_{2} - h_{1} ) x } - 1 \right\}  
\nonumber \\
&-&
\epsilon c_{12} s_{12} s_{ (\phi - \theta_{13}) } 
\left( c_{23} s_{\phi} s^2_\psi e^{ i \delta} - s_{23} \cos 2\phi c_{\psi} s_\psi 
\right) 
\frac{ \Delta_{ \text{ren} } }{ h_{3} - h_{1} } 
\left\{ e^{ i ( h_{2} - h_{1} ) x } - 1 \right\} 
\nonumber \\
&+& 
\epsilon c_{12} s_{12} s_{ (\phi - \theta_{13}) } 
\left( 
c_{23} s_{\phi} c^2_\psi e^{ i \delta} + s_{23} \cos 2\phi c_{\psi} s_\psi 
\right) 
\frac{ \Delta_{ \text{ren} } }{ h_{3} - h_{2} } 
\left\{ e^{ - i ( h_{3} - h_{2} ) x } - 1 \right\} 
\nonumber \\
&+& 
\epsilon c_{12} s_{12} s_{ (\phi - \theta_{13}) } 
\left( 
c_{23} s_{\phi} s^2_\psi e^{ i \delta} - s_{23} \cos 2\phi c_{\psi} s_\psi 
\right) 
\frac{ \Delta_{ \text{ren} } }{ h_{3} - h_{1} } 
\left\{ e^{ - i ( h_{3} - h_{2} ) x } - 1 \right\}. 
\label{S-emu-Reph2}
\end{eqnarray} 

Is the physical interpretation of $S_{\alpha \beta}^{ \text{Reph-1} }$ and $S_{\alpha \beta}^{ \text{Reph-2} }$ available? 
If we focus on the leading order terms which would give the atmospheric-scale enhancement, each one of $S_{\alpha \beta}^{ \text{Reph-1} }$ and $S_{\alpha \beta}^{ \text{Reph-2} }$ has the factor 
$\left[ e^{ - i ( \lambda_{3} - \lambda_{1} ) x / 2E } - 1 \right]$, or 
$\left[ e^{ - i ( \lambda_{3} - \lambda_{2} ) x / 2E } - 1 \right]$. 
They lead to the form of the probability proportional to $\sin^2 \frac{ ( \lambda_{3} - \lambda_{1} ) x }{ 4E } $ and $\sin^2 \frac{ ( \lambda_{3} - \lambda_{2} ) x }{ 4E } $, respectively, which means the 1-3 and 2-3 level crossings around the atmospheric-scale resonance. The feature of the leading term would suggest that $S_{\alpha \beta}^{ \text{Reph-1} }$ and $S_{\alpha \beta}^{ \text{Reph-2} }$ describe the inverted mass ordering (IMO) and normal mass ordering (NMO), respectively.\footnote{
%%%%%%%%%%%%% footnote %%%%%%%%%%%%%%%
In the NMO the atmospheric level crossing is between the eigenvalues $\lambda_{3} > \lambda_{2}$, and in the IMO between $\lambda_{1} > \lambda_{3}$. See Fig.~1 in ref.~\cite{Denton:2016wmg}. }
Of course, this interpretation cannot be applied literally to the whole amplitudes, because they are related by the phase factor, and the observable probabilities are exactly the same. Therefore, an intuitive interpretation exists which is valid for the leading atmospheric terms, but it does not prevail to the whole amplitudes. 

\section{$S$ matrix rephasing invariance realized locally by the reparametrization (Rep) symmetry}
\label{sec:analytic-mapping}

We have defined in eq.~\eqref{S-Reph-12-def} the two differently rephased amplitudes $S_{\alpha \beta}^{ \text{Reph-1} }$ and $S_{\alpha \beta}^{ \text{Reph-2} }$, and given their explicit expressions in the $\nu_{\mu} \rightarrow \nu_{e}$ channel~$(\alpha \beta = e \mu)$, eqs.~\eqref{S-emu-Reph1} and \eqref{S-emu-Reph2}. We note that if they are just written down, their exact equivalence in the observable is not so obvious. It would be illuminating if physical equivalence between $S_{\alpha \beta}^{ \text{Reph-1} }$ and $S_{\alpha \beta}^{ \text{Reph-2} }$ can be revealed explicitly by an analytic mapping between them as in eq.~\eqref{Phi-transf}. We will show below that this is indeed what happens, and the mapping is executed by the Rep symmetry transformations obtained in ref.~\cite{Minakata:2021dqh}, see Table~\ref{tab:DMP-symmetry}. This is an entirely new way of demonstration of the rephasing invariance to our knowledge. 

\subsection{Mapping between $S_{\alpha \beta}^{ \text{Reph-1} }$ and $S_{\alpha \beta}^{ \text{Reph-2} }$ in the $\nu_{\mu} \rightarrow \nu_{e}$ and $\nu_{\mu} \rightarrow \nu_{\tau}$ channels}
\label{sec:mapping-mue-mutau} 

One can readily show that $S_{e \mu}^{ \text{Reph-2} }$ is mapped to $S_{e \mu}^{ \text{Reph-1} }$ by the transformations of Symmetry X-DMP as 
\begin{eqnarray} 
&& 
S_{\alpha \beta}^{ \text{Reph-2} } 
\rightarrow 
\pm S_{\alpha \beta}^{ \text{Reph-1} }, 
\label{mapping-emu}
\end{eqnarray} 
and vice versa. Here $\alpha \beta = e \mu$. See eqs.~\eqref{S-emu-Reph2} and~\eqref{S-emu-Reph1} for their explicit forms, and the transformations of all Symmetry X (X=IA, IB, $\cdot \cdot \cdot $, IVB) are given in Table~\ref{tab:DMP-symmetry}. This is what we mean by ``the $S$ matrix rephasing invariance represented locally by the Rep symmetry''.\footnote{
%%%%%%%%%%%%% footnote %%%%%%%%%%%%%%%%
A historical note: Symmetry IA-DMP in our unified nomenclature~\cite{Minakata:2021dqh,Minakata:2022yvs} is known in the original DMP paper~\cite{Denton:2016wmg}. But, the author was surprised by the reappearance of IA-DMP in a place totally unexpected, the investigation of neutrino amplitude decomposition in matter~\cite{Minakata:2020oxb}. We wrote in section VE in ref.~\cite{Minakata:2020oxb} that: ``Thus, we have identified the origin of the $\psi$ symmetry: It is due to the freedom of doing rephasing in the $S$ matrix, assuming the probabilistic nature of quantum mechanics.'' Our present treatment in this paper generalizes the same statement to all the eight Rep symmetries in DMP. } 

We have used the generic flavor index in eq.~\eqref{mapping-emu} because one can show that it holds also in the $\nu_{\mu} \rightarrow \nu_{\tau}$ channel, $\alpha \beta = \tau \mu$. This task can be carried out straightforwardly by doing the Rep symmetry transformations in Table~\ref{tab:DMP-symmetry} onto one of the expressions of $S_{\tau \mu}^{ \text{Reph-j} }$ ($j=1,2$) given in Appendix~\ref{sec:S-tau-mu}, see eqs.~\eqref{S-taumu-Reph1-SM} and ~\eqref{S-taumu-Reph2-SM}. 
Interestingly, the $\pm$ sign dependence on symmetry type X in the $\nu_{\mu} \rightarrow \nu_{\tau}$ channel is different from the sign dependence in the $\nu_{\mu} \rightarrow \nu_{e}$ channel. In the $\nu_{\mu} \rightarrow \nu_{e}$ channel, the plus sign is for X = I and IV, and the minus sign is for X = II, and III. Whereas in the $\nu_{\mu} \rightarrow \nu_{\tau}$ channel, the plus sign is for X = I and III, and the minus sign is for X = II, and IV. 

We remark that here and in what follows if we make the statement like above, ``the plus sign is for X = I and IV'', we mean that the statement holds for the both A type (without $\delta$ transformation) and B (with $\delta$) type, i.e., X=IA and IB, and IVA and IVB. We also use the phrase ``transforms under Symmetry X'' as our shorthand expression of ``transforms under the transformations which belong to Symmetry X''.

%%%%%%%%%%%%%%%%%%%%%%%%%%%%%%%%%%%%%%%
\begin{table}[h!]
\vglue 0.2cm
\begin{center}
\caption{The eight reparametrization (Rep) symmetries of the $1 \leftrightarrow 2$ state exchange type in DMP. The nomenclature of the symmetry Symmetry X (X = IA, IB, IIA, IIB, IIIA, IIIB, IVA, and IVB) is established in ref.~\cite{Minakata:2021dqh}, from where this table is taken. The notations are such that $\lambda_{j}$ ($j=1,2$) are the first two eigenvalues of $2E H$, $\psi$ and $\phi$ denote $\theta_{12}$ and $\theta_{13}$ in matter, respectively. 
}
\label{tab:DMP-symmetry}
\vglue 0.2cm
\begin{tabular}{c|c|c}
\hline 
Symmetry & 
Vacuum parameter transformations & 
Matter parameter transformations
\\
\hline 
\hline 
Symmetry IA & 
none & 
$\lambda_{1} \leftrightarrow \lambda_{2}$, 
$c_{\psi} \rightarrow \mp s_{\psi}$, 
$s_{\psi} \rightarrow \pm c_{\psi}$. \\
\hline 
Symmetry IB & 
$\theta_{12} \rightarrow - \theta_{12}$, 
$\delta \rightarrow \delta + \pi$. & 
$\lambda_{1} \leftrightarrow \lambda_{2}$, 
$c_{\psi} \rightarrow \pm s_{\psi}$, 
$s_{\psi} \rightarrow \pm c_{\psi}$. \\
\hline
Symmetry IIA & 
$\theta_{23} \rightarrow - \theta_{23}$, 
$\theta_{12} \rightarrow - \theta_{12}$. & 
$\lambda_{1} \leftrightarrow \lambda_{2}$, 
$c_{\psi} \rightarrow \pm s_{\psi}$, 
$s_{\psi} \rightarrow \pm c_{\psi}$. \\
\hline 
Symmetry IIB & 
$\theta_{23} \rightarrow - \theta_{23}$, 
$\delta \rightarrow \delta + \pi$. & 
$\lambda_{1} \leftrightarrow \lambda_{2}$, 
$c_{\psi} \rightarrow \mp s_{\psi}$, 
$s_{\psi} \rightarrow \pm c_{\psi}$. \\
\hline 
Symmetry IIIA & 
$\theta_{13} \rightarrow - \theta_{13}$, 
$\theta_{12} \rightarrow - \theta_{12}$. & 
$\lambda_{1} \leftrightarrow \lambda_{2}$, 
$\phi \rightarrow - \phi$, \\ 
 & & 
$c_{\psi} \rightarrow \pm s_{\psi}$, 
$s_{\psi} \rightarrow \pm c_{\psi}$ \\
\hline 
Symmetry IIIB & 
$\theta_{13} \rightarrow - \theta_{13}$, 
$\delta \rightarrow \delta + \pi$. & 
$\lambda_{1} \leftrightarrow \lambda_{2}$, 
$\phi \rightarrow - \phi$, \\ 
 & & 
$c_{\psi} \rightarrow \mp s_{\psi}$, 
$s_{\psi} \rightarrow \pm c_{\psi}$. \\
\hline 
Symmetry IVA & 
$\theta_{23} \rightarrow - \theta_{23}$, 
$\theta_{13} \rightarrow - \theta_{13}$. & 
$\lambda_{1} \leftrightarrow \lambda_{2}$, 
$\phi \rightarrow - \phi$, \\ 
 & & 
$c_{\psi} \rightarrow \mp s_{\psi}$, 
$s_{\psi} \rightarrow \pm c_{\psi}$. \\
\hline 
Symmetry IVB & 
$\theta_{23} \rightarrow - \theta_{23}$, 
$\theta_{13} \rightarrow - \theta_{13}$, & 
$\lambda_{1} \leftrightarrow \lambda_{2}$, 
$\phi \rightarrow - \phi$, \\ 
 &
$\theta_{12} \rightarrow - \theta_{12}$, $\delta \rightarrow \delta + \pi$. 
 &
$c_{\psi} \rightarrow \pm s_{\psi}$, $s_{\psi} \rightarrow \pm c_{\psi}$. \\
\hline 
\end{tabular}
\end{center}
\vglue -0.4cm 
\end{table}
%%%%%%%%%%%%%%%%%%%%%%%%%%%%%%%%%%%%%

We have shown in this section that the rephasing invariance is realized as the local mapping between $S_{\alpha \beta}^{ \text{Reph-1} }$ and $S_{\alpha \beta}^{ \text{Reph-2} }$ induced by the Rep symmetry transformations. One of the remaining questions is in which way the $\pm$ sign in eq.~\eqref{mapping-emu} is determined and why the sign is not completely identical (opposite in Symmetry III and IV) between the $\nu_{\mu} \rightarrow \nu_{e}$ and the $\nu_{\mu} \rightarrow \nu_{\tau}$ channels. These problems will be settled in section~\ref{sec:transf-S}. 
Extension of our discussion to the remaining flavor channels, including the disappearance channels, can be done straightforwardly. But we will do it in a somewhat different way in sections~\ref{sec:rephasing-from-Rep} and~\ref{sec:all-orders}. 

\section{The Rep symmetry implies the $S$ matrix rephasing invariance} 
\label{sec:rephasing-from-Rep} 

Once we recognize all the Rep symmetries and their role in the context of the $S$ matrix rephasing invariance, we can reverse the argument. That is, the Rep symmetry implies the $S$ matrix rephasing invariance. 
Or, more precisely speaking, once a Rep symmetry of the state exchange type is given, one can create a one-parameter family of the two differently rephased $S$ matrices with the same physical contents. 

Let $S_{\alpha \beta}$ be one of the flavor basis $S$ matrix elements, and assume that it transforms under Symmetry X-DMP as 
\begin{eqnarray} 
&&
S_{\alpha \beta} \rightarrow \pm S_{\alpha \beta}. 
\label{S-transf} 
\end{eqnarray} 
In fact, the $S$ matrix elements $S_{e \mu}$ and $S_{\tau \mu}$ computed to the first order in perturbation theory, as given in eqs.~\eqref{S-emu} and~\eqref{S-taumu-SM} respectively, satisfy eq.~\eqref{S-transf} with the same (channel dependent) $\pm$ signs as in eq.~\eqref{mapping-emu}. 
Let us next introduce an arbitrary real parameter $\xi$ and define a family of the two rephased $S$ matrices parametrized by $\xi$:
\begin{eqnarray} 
&&
S_{\alpha \beta}^{ \text{Reph-1} } (\xi)
\equiv 
e^{ i \xi \frac{ \lambda_{1} }{2E} x} S_{\alpha \beta}, 
\hspace{10mm} 
S_{\alpha \beta}^{ \text{Reph-2} } (\xi) 
\equiv 
e^{ i \xi \frac{ \lambda_{2} }{2E} x} S_{\alpha \beta}.
\label{Reph-12-family} 
\end{eqnarray} 
We are aware that the functional form of the phase factors can be more general, such as $e^{ i F ( \lambda_{i} x / 2E ) }$ where $F(y)$ is a function of the real variable $y$. But, we do not try to elaborate this point because the one-parameter family defined in eq.~\eqref{Reph-12-family} is sufficient for our purpose. 

Then, it is obvious that under Symmetry X-DMP  
\begin{eqnarray} 
&& 
S_{\alpha \beta}^{ \text{Reph-2} } (\xi) 
\rightarrow 
\pm S_{\alpha \beta}^{ \text{Reph-1} } (\xi) 
\label{mapping-alpha-beta}
\end{eqnarray}
given that the state exchange $\lambda_{1} \leftrightarrow \lambda_{2}$ is involved in all the Symmetry X. See Table~\ref{tab:DMP-symmetry}. Therefore, a Rep symmetry of the state exchange type implies a $S$ matrix rephasing invariance between $S_{\alpha \beta}^{ \text{Reph-1} } (\xi)$ and $S_{\alpha \beta}^{ \text{Reph-2} } (\xi)$. As far as the Rep symmetry is known for all the flavor channels, which is the case in our present discussion, the above procedure applies to the all oscillation channels. 
Existence of a $\xi$ parametrized family of the rephased amplitudes suggests that the rephasing invariance is larger than the Rep symmetry. 

It appears that this path of showing the existence of the $S$ matrix rephasing invariance starting from the Rep symmetry is more promising to prove the rephasing invariance to all orders in the DMP perturbation theory. Remember that our foregoing treatment in sections~\ref{sec:analytic-mapping} and~\ref{sec:rephasing-from-Rep} is valid only to first order. In section~\ref{sec:all-orders} we will pursue this path toward the all-order proof of the rephasing invariance, but without a complete success. 

\section{$S$ matrix rephasing invariance does not necessarily imply the Rep symmetry}
\label{sec:rephasing-wider}

While our foregoing discussions might have left an impression that a $S$ matrix rephasing invariance and a Rep symmetry of the state exchange type are essentially equivalent to each other, apart from the degeneracy parametrized by $\xi$, it is not true. It appears that the $S$ matrix rephasing invariance, in general, may be much larger than the Rep symmetry. However, speaking strictly and frankly, our knowledge on the relationship between them is very limited, and we still have many questions, including: 
\begin{enumerate} 

\item
For a given $S$ matrix rephasing invariance, even restricting to the type we have discussed in sections~\ref{sec:analytic-mapping} and~\ref{sec:rephasing-from-Rep}, does a Rep symmetry always exist to realize it locally? 

\item
If yes, how can the Rep symmetry be constructed? If not, what is the criterion for finding the Rep symmetry for a local realization of $S$ matrix rephasing invariance?

\end{enumerate}
Though we are not able to fully address these questions in this paper, we present below a simple-minded example which provides a negative answer to the first question. 

\subsection{Absence of the 1-3 state exchange symmetry in DMP} 
\label{sec:no-13-DMP} 

We argue below that the physical equivalence between $S_{e \mu}^{ \text{Reph-1} }$ and $S_{e \mu}^{ \text{Reph-3} } \equiv e^{ i ( \lambda_{3} / 2E ) x} S_{e \mu}$ cannot be represented locally by the 1-3 exchange Rep symmetry in DMP.\footnote{
%%%%%%%%%%%% footnote %%%%%%%%%%%%
This terminology of ``1-3 exchange symmetry'' is of a symbolic nature. It actually means the 1-3 (2-3) exchange symmetry in IMO (NMO). } 
To investigate this problem we calculate the $e^{ i ( \lambda_{3} / 2E ) x}$-rephased amplitude for the given $S$ matrix element in eq.~\eqref{S-emu}:  
\begin{eqnarray} 
\hspace{-8mm}
S_{e \mu}^{ \text{Reph-3} } 
&\equiv &
e^{ i \frac{ \lambda_{3} }{2E} x} S_{e \mu} 
%\nonumber \\&=& 
= c_{\phi} \left( c_{23} c_\psi s_\psi e^{ i \delta} - s_{23} s_{\phi} s^2_\psi \right) 
\left\{ e^{ i ( h_{3} - h_{2} ) x} - 1 \right\} 
\nonumber \\
&-& 
c_{\phi} \left( c_{23} c_\psi s_\psi e^{ i \delta} + s_{23} s_{\phi} c^2_\psi \right) 
\left\{ e^{ i ( h_{3} - h_{1} ) x} - 1 \right\} 
\nonumber \\
&-& 
\epsilon c_{12} s_{12} s_{ (\phi - \theta_{13}) } 
\left( 
c_{23} s_{\phi} c^2_\psi e^{ i \delta} + s_{23} \cos 2\phi c_{\psi} s_\psi 
\right) 
\frac{ \Delta_{ \text{ren} } }{ h_{3} - h_{2} } 
\left\{ e^{ i ( h_{3} - h_{2} ) x} - 1 \right\} 
\nonumber \\
&-& 
\epsilon c_{12} s_{12} s_{ (\phi - \theta_{13}) } 
\left( 
c_{23} s_{\phi} s^2_\psi e^{ i \delta} - s_{23} \cos 2\phi c_{\psi} s_\psi 
\right) 
\frac{ \Delta_{ \text{ren} } }{ h_{3} - h_{1} } 
\left\{ e^{ i ( h_{3} - h_{1} ) x} - 1 \right\}, 
\label{S-emu-Reph3}
\end{eqnarray}
where $h_{k} \equiv \lambda_{k} / 2E$. 
Differing only by the phase factors, $S_{e \mu}^{ \text{Reph-3} }$ must be physically equivalent to $S_{e \mu}^{ \text{Reph-1} }$, or to $S_{e \mu}^{ \text{Reph-2} }$, given respectively in eqs.~\eqref{S-emu-Reph1} and \eqref{S-emu-Reph2}. 

If the equivalence is shown manifestly by the presence of the Rep symmetry, $S_{e \mu}^{ \text{Reph-3} }$ must turn to $\pm S_{e \mu}^{ \text{Reph-1} }$ under the transformations $h_{3} \rightarrow h_{1} $, $h_{1} \rightarrow h_{3}$, and $c_{\phi} \rightarrow c_{\phi}^{\prime}$, $s_{\psi} \rightarrow s_{\psi}^{\prime}$ etc. However, the resultant expression obtained by these transformations reads 
\begin{eqnarray} 
\hspace{-10mm}
S_{e \mu}^{ \text{Reph-3} } 
&\rightarrow& 
c_{\phi}^{\prime} \left\{ c_{23}^{\prime} c_\psi ^{\prime} s_\psi ^{\prime} e^{ i \delta^{\prime}} - s_{23}^{\prime} s_{\phi}^{\prime} ( s_\psi ^{\prime} )^2 \right\} 
\left\{ e^{ i ( h_{1} - h_{2} ) x} - 1 \right\} 
\nonumber \\
&-& 
c_{\phi}^{\prime} \left\{ c_{23}^{\prime} c_\psi^{\prime} s_\psi^{\prime} e^{ i \delta^{\prime} } + s_{23}^{\prime} s_{\phi}^{\prime} ( c_\psi ^{\prime} )^2 \right\} 
\left\{ e^{ i ( h_{1} - h_{3} ) x} - 1 \right\} 
\nonumber \\
&-& 
\epsilon c_{12}^{\prime} s_{12}^{\prime} s_{ (\phi^{\prime} - \theta_{13}^{\prime}) } 
\left\{ 
c_{23}^{\prime} s_{\phi}^{\prime} ( c_\psi ^{\prime} )^2 e^{ i \delta^{\prime} } 
+ s_{23}^{\prime} \cos 2\phi^{\prime} c_{\psi}^{\prime} s_\psi ^{\prime}
\right\} 
\frac{ \Delta_{ \text{ren} } }{ h_{1} - h_{2} } 
\left\{ e^{ i ( h_{1} - h_{2} ) x} - 1 \right\} 
\nonumber \\
&-& 
\epsilon c_{12}^{\prime} s_{12}^{\prime} s_{ (\phi^{\prime} - \theta_{13}^{\prime}) } 
\left\{ 
c_{23}^{\prime} s_{\phi}^{\prime} ( s_\psi ^{\prime} )^2 e^{ i \delta^{\prime} } 
- s_{23}^{\prime} \cos 2\phi^{\prime} c_{\psi}^{\prime} s_\psi ^{\prime} 
\right\} 
\frac{ \Delta_{ \text{ren} } }{ h_{1} - h_{3} } 
\left\{ e^{ i ( h_{1} - h_{3} ) x} - 1 \right\}. 
\label{S-emu-Reph3-transf}
\end{eqnarray}
Clearly the third line in eq.~\eqref{S-emu-Reph3-transf}, which is proportional to 
$\frac{ \Delta_{ \text{ren} } }{ h_{2} - h_{1} } \left\{ e^{ - i ( h_{2} - h_{1} ) x} - 1 \right\}$, does not find a counterpart in $S_{e \mu}^{ \text{Reph-1} }$ in eq.~\eqref{S-emu-Reph1}. Notice that we are talking about the kinematical factor only, and hence the statement holds irrespective of the actual transformation properties of $\theta_{ij}$, $\psi$, and $\phi$ etc. Therefore, it is shown that any analytic mapping based on the eigenvalue exchange $h_{1} \leftrightarrow h_{3}$ to guarantee the equivalence between $S_{e \mu}^{ \text{Reph-3} }$ and $S_{e \mu}^{ \text{Reph-1} }$ does not exist. If we repeat the same exercise we reach the same conclusion of no $S_{e \mu}^{ \text{Reph-3} }$ - $S_{e \mu}^{ \text{Reph-2} }$ mapping to guarantee the equivalence between them. 

Thus, we conclude that physical equivalence between $S_{e \mu}^{ \text{Reph-3} }$ and $S_{e \mu}^{ \text{Reph-1} }$, or between $S_{e \mu}^{ \text{Reph-3} }$ and $S_{e \mu}^{ \text{Reph-2} }$, cannot be represented by the existence of a Rep symmetry, as far as we restrict to the simple two-state exchange type. No pure 1-3, or 2-3, state exchange Rep symmetry exists in DMP to serve for the local realization of the rephasing invariance. 

\subsection{``One-resonance - one-symmetry'' picture: Did it fail?}
\label{sec:one-res-one-sym} 

In a previous paper~\cite{Minakata:2022pyr} we have argued that the 1-3 state exchange symmetry must exist in DMP, which is in apparent contradiction to the above result. We ask ourselves which one is correct, and if possible, we would like to understand the reason for the two differing statements to draw some implications. 

In ref.~\cite{Minakata:2022yvs} we have presented the ``one-resonance - one-symmetry'' picture to argue that the Rep symmetry of the $i$-$j$ state exchange type exists at around the resonance associated with the $i$-$j$ level crossing.\footnote{
%%%%%%%%%%%%% footnote %%%%%%%%%%%%%%
We admit that this sentence may be confusing even though it was used to define the one-resonance - one-symmetry picture. In locally-valid theories (see below) it may be legitimate as the regions of validity of SRP and helio-perturbation theories are restricted to around the solar and the atmospheric-scale enhancements, respectively. But in DMP the 1-2 exchange symmetry exists in the whole region where the perturbation theory is defined, not just around the solar resonance. The statement on the picture is intended to illuminate which type of the symmetry exists in a given theory, and not on where the symmetry resides. }
Our basic motivation is to understand the nature of the Rep symmetry, whether it is merely a framework-dependent entity, or a reflection of the physical property of neutrino evolution in matter. In consistent with this picture we have found the essentially identical 1-2 state exchange symmetries~\cite{Minakata:2022pyr}  both in the locally-valid SRP (solar-resonance perturbation) theory~\cite{Martinez-Soler:2019nhb} and in the globally-valid DMP~\cite{Denton:2016wmg}. In simple terms, the locally-valid theory~\cite{Arafune:1996bt,Cervera:2000kp,Freund:2001pn,Akhmedov:2004ny,Minakata:2015gra,Martinez-Soler:2019nhb} can describe the region around one of the solar-scale or the atmospheric-scale resonances, and the globally-valid one~\cite{Agarwalla:2013tza,Denton:2016wmg} the both. See refs.~\cite{Minakata:2022yvs,Minakata:2022pyr} for more about the concepts. We have stated there as follows: Since the 1-3 exchange symmetry is known to exist~\cite{Minakata:2021goi} in the locally-valid helio-perturbation theory~\cite{Minakata:2015gra} it must exists also in the globally-valid DMP. However, we have just shown that the pure 1-3 (or 2-3) state exchange symmetry does not exist in DMP, assuming the function of the Rep symmetry as locally realizing the $S$ matrix rephasing invariance. 

The absence of the pure 2-3 (NMO) or 1-3 (IMO) state exchange symmetry in DMP suggests that the effect of the third state (state 1 in NMO, and state 2 in IMO) cannot be ignored. Then, the question is whether this is a reasonable thing to expect, or not, which we answer in the positive as in below. It is known that the atmospheric resonance is wider than the solar one in the energy region affected. Let us tentatively define the atmospheric-resonance-affected region by $\phi$, the matter-affected $\theta_{13}$, as $\frac{\pi}{4} - \frac{\pi}{8} \lsim \phi \lsim \frac{\pi}{4} + \frac{\pi}{8}$. It corresponds, roughly speaking, to the energy - matter-density region 10 (g/cm$^3$) GeV $\lsim Y_{e} \rho E \lsim$ 20 (g/cm$^3$) GeV, where $\rho$ and $Y_e$ denote, respectively, the matter density and number of electrons per nucleon in matter. See Fig.~1 in ref.~\cite{Denton:2016wmg}. In this region $\psi$, the matter-affected $\theta_{12}$, changes albeit by small amount, signaling small perturbation by the solar-scale enhancement in the region of the atmospheric resonance. For antineutrinos $Y_{e} \rho E$ flips sign. 
In sharp contrast to the atmospheric resonance, the similarly defined solar-resonance-affected region of $\psi$ is very narrow, at least an order of magnitude narrower than the atmospheric $\phi$ region. In the solar resonance region, $\phi$ essentially freezes to its vacuum value $\theta_{13}$, indicating negligible perturbation by the atmospheric resonance. This is consistent with the formulation of SRP~\cite{Martinez-Soler:2019nhb} in which $\theta_{13}$ stays at its vacuum value, not elevated to the matter-affected one. 

To summarize, the state-exchange Rep symmetry around the atmospheric resonance would not exist as a pure 1-3 (or 2-3) exchange type, but the one perturbed by the third state in DMP. While the single, separated resonance picture is valid for the solar resonance (1-2 level crossing), it does not for the atmospheric resonance (2-3, or 1-3 level crossing). We believe it the reason for missing the pure 1-3 (or 2-3) exchange symmetry in DMP. It is indeed interesting that the Rep symmetry can reveal how well a resonance can be treated as an isolated object, or perturbed by the other resonant enhancement. 

Absence of the pure 1-3 (or 2-3) exchange symmetry in DMP seems to be consistent with the result of our preliminary investigation of the 1-3 state exchange Rep symmetry using the SF formalism. Assuming that all these reasonings are correct, we must find the Rep symmetry of the extended three states exchange type to completely settle this issue. It may be related to the conjecture on possible further enlargement of the Rep symmetry, as presented in ref.~\cite{Minakata:2022zua}. 

\section{Rep symmetry in neutrino oscillations: What is it?} 
\label{sec:Rep-what-is} 

While we intend to address the problem of extending our discussion to all orders in DMP perturbation theory in the next section~\ref{sec:all-orders}, which is slightly technical, it may be appropriate to give the summary statement on the question ``What is the nature of the Rep symmetry?''. We try to give our best answer allowed by our current understanding. 

We make here the two statements on the Rep symmetry. The key descriptions most relevant for therm are given in sections~\ref{sec:analytic-mapping} and \ref{sec:rephasing-wider}, respectively, in order. 
\begin{itemize}

\item 
The Rep symmetry offers a local and manifest realization of the $S$ matrix rephasing invariance. We interpret this feature as that the Rep symmetry is originated in quantum mechanics of the neutrino oscillation. To our knowledge, the characterization of the symmetry and such a way of realizing the rephasing invariance are entirely new. 

\item 
The Rep symmetry provides an independent probe into the theory of neutrino oscillation. We have argued in section~\ref{sec:one-res-one-sym} that absence of the pure 1-3 state exchange symmetry in DMP indicates that description of the atmospheric-scale resonance as isolated from the solar resonance cannot be perfect. 

\end{itemize}

Now, the neutrino oscillation and flavor conversion are the indispensable ingredients of the $\nu$SM, even though the mass generation mechanism for neutrinos is not built in it. Necessity of deeper theoretical understanding of these phenomena cannot be over-emphasized, in recalling the enormous efforts devoted to found the theoretical basis of the SM in the seventies and eighties. The SF search for the Rep symmetry has been motivated along the line of thought. If our view of the Rep symmetry coming from quantum mechanics is supported,  symmetry violation may involve violation of one of the principles of quantum mechanics. This is too deep a subject to enter in the present paper. 

Though the interpretation of our second statement above must be done with great care, it is interesting that the Rep symmetry can trigger the question of whether a given perturbative framework is sound or not. We will try to address this issue in section~\ref{sec:consistency}. 

\section{Toward local realization of the $S$ matrix rephasing invariance to all orders}
\label{sec:all-orders} 

In sections~\ref{sec:analytic-mapping} and~\ref{sec:rephasing-from-Rep} we have revealed the intimate relationships between the $S$ matrix rephasing invariance and the Rep symmetry. The arguments proceeded along the two opposite directions: 
(1) If we have the two rephased $S$ matrices $S_{\alpha \beta}^{ \text{Reph-1} }$ and $S_{\alpha \beta}^{ \text{Reph-2} }$ as in eqs.~\eqref{S-emu-Reph1} and \eqref{S-emu-Reph2}, respectively, we can show that there exist the local transformations of the dynamical variables which map $S_{\alpha \beta}^{ \text{Reph-1} }$ to $S_{\alpha \beta}^{ \text{Reph-2} }$, or vice versa, realizing manifestly the $S$ matrix rephasing invariance. 
(2) Instead, if we know that a Rep symmetry of the 1-2 state exchange type exists in the system, we can construct the two rephased $S$ matrices using the $\xi$ parameter extension $S_{\alpha \beta}^{ \text{Reph-1} } (\xi) \equiv e^{ i \xi (\lambda_{1}/2E) x} S_{\alpha \beta}$ and $S_{\alpha \beta}^{ \text{Reph-2} } (\xi) \equiv e^{ i \xi (\lambda_{2}/2E) x} S_{\alpha \beta}$ by which their physical equivalence can be proven. But, our treatment at hand for the arguments (1) and (2) is valid to the first order in the DMP perturbation theory. 

Is an all-order treatment possible? Proceeding via the first path (1) is hard to achieve because an explicit computation of the $S$ matrix to all orders is beyond our current technology.\footnote{
%%%%%%%%%%%%% footnote %%%%%%%%%%%%%%%
We leave an interesting second-order exercise to those who are interested to carry out. } 
The second path (2) may be divided into the two parts (2a) and (2b): 
(2a) is to obtain the Rep symmetry valid to all orders in the DMP perturbation theory. 
(2b) is to obtain the eigenvalues $\Lambda_{1}$ and $\Lambda_{2}$, all-order versions of $\lambda_{1}$ and $\lambda_{2}$, and verify the exchange property $\Lambda_{1} \leftrightarrow \Lambda_{2}$ under the transformations of Symmetry X for all X=IA, IB, $\cdot \cdot \cdot $, IVB. 
Once the both (2a) and (2b) are completed our goal is fulfilled: 
$S_{\alpha \beta}^{ \text{Reph-1} } (\xi) \equiv e^{ i \xi (\Lambda_{1}/2E) x} S_{\alpha \beta}$ and $S_{\alpha \beta}^{ \text{Reph-2} } (\xi) \equiv e^{ i \xi (\Lambda_{2}/2E) x} S_{\alpha \beta}$ are physically equivalent to each other. We will show that (2a) can be pursuit and offers an explanation of the $\pm$ sign in eqs.~\eqref{mapping-emu} and~\eqref{mapping-alpha-beta}, which are partly different between the $\nu_{\mu} \rightarrow \nu_{e}$ and the $\nu_{\mu} \rightarrow \nu_{\tau}$ channels. We must admit that (2b) eludes us at this moment, but we will show that it can be done to the fourth order.

Since we have the Hamiltonian proof of the Rep symmetry in DMP~\cite{Minakata:2021dqh}, as well as in the other perturbative frameworks~\cite{Minakata:2021goi,Minakata:2022zua,Minakata:2022yvs}, which is valid to all orders in perturbation theory, we should be able to achieve the goal (2a). Existence of the perturbative proof of the Rep symmetry to all orders strongly suggests that the exchange property of the all-order eigenvalues $\Lambda_{1} \leftrightarrow \Lambda_{2}$ holds under Symmetry X, but it lacks an explicit proof. 

\subsection{The DMP perturbation theory: A minimal recollection} 
\label{sec:recollection-DMP} 

To pursuit our goal (2a) we need a minimal knowledge of the DMP perturbation theory. The standard three neutrino evolution in matter is described by the Schr\"odinger equation $i \frac{d}{dx} \check{\nu} = \check{H} \check{\nu}$ in the vacuum mass eigenstate basis, or the check basis, with the Hamiltonian 
$\check{H} = \frac{1}{2E} \left[ \text{diag} (0, \Delta m^2_{21}, \Delta m^2_{31} ) + U^{\dagger} \text{diag} (a, 0, 0) U \right]$, 
where $\check{\nu}$ is related with the flavor eigenstate as $\nu = U \check{\nu}$ by using the flavor mixing matrix $U$ in eq.~\eqref{U-SOL-def}. $a(x) = 2 \sqrt{2} G_F n_e E$ denotes the Wolfenstein matter potentials~\cite{Wolfenstein:1977ue} due to charged current reactions. $G_F$ is the Fermi constant, and $n_e$ and $n_n$ the electron and neutron number densities in matter, respectively. 

Starting from $\check{H}$ the diagonalization of the Hamiltonian can be performed to produce the approximately diagonal Hamiltonian $\bar{H}$ as~\cite{Denton:2016wmg} 
\begin{eqnarray} 
\bar{H} 
&=&
U_{12}^{\dagger} (\psi, \delta) U^\dagger_{13}(\phi) 
U_{13} (\theta_{13}) U_{12} (\theta_{12}) 
\check{H} 
U_{12} (\theta_{12})^{\dagger} U_{13} (\theta_{13}) ^{\dagger} 
U_{13} (\phi) U_{12} (\psi, \delta), 
\label{barH-def} 
\end{eqnarray}
where $U_{12} (\psi, \delta)$ denotes $U_{12} (\theta_{12}, \delta)$ with $\theta_{12}$ replaced by $\psi$, for example. The explicit forms of the rotation matrices $U_{ij}$ are given in eq.~\eqref{U-SOL-def}. The bar-basis Hamiltonian is given by 
\begin{eqnarray} 
\bar{H} 
&=&
\frac{1}{2E} 
\left[
\begin{array}{ccc}
\lambda_{1} & 0 & 0 \\
0 & \lambda_{2} & 0 \\
0 & 0 & \lambda_{3} \\
\end{array}
\right] 
+
\epsilon c_{12} s_{12} 
\sin ( \phi - \theta_{13} ) \frac{\Delta m^2_{ \text{ren} }}{2E} 
\left[
\begin{array}{ccc}
0 & 0 & - s_\psi \\
0 & 0 & c_\psi e^{ - i \delta} \\
- s_\psi & c_\psi e^{ i \delta} & 0 
\end{array}
\right]. 
\label{barH-0th-1st}
\end{eqnarray}
The flavor basis Hamiltonian $H_{ \text{flavor} }$ and the $S_{ \text{flavor} }$ matrix are related to the bar-basis ones, 
\begin{eqnarray} 
H_{ \text{flavor} } 
&=&  
U_{23} U_{13}(\phi) U_{12} (\psi, \delta) 
\bar{H} 
U_{12}^{\dagger} (\psi, \delta) U^\dagger_{13}(\phi) U_{23}^{\dagger}, 
\nonumber \\
S_{ \text{flavor} } 
&=& 
U_{23} U_{13}(\phi) U_{12} (\psi, \delta) 
\bar{S} 
U_{12}^{\dagger} (\psi, \delta) U^\dagger_{13} (\phi) U_{23}^{\dagger}. 
\label{H-S-flavor-bar}
\end{eqnarray}
Hereafter $U_{23}$ without argument implies the 2-3 rotation matrix in vacuum, $U_{23} (\theta_{23})$. 

To understand better the DMP formalism in the context of this paper the author recommends to visit ref.~\cite{Minakata:2022zua}. It develops an improved method for proving the Hamiltonian invariance, which facilitates invariance proof of the $S$ matrix in what follows.  One can eliminate the pieces related to unitarity violation (UV) in the description in ref.~\cite{Minakata:2022zua} to see the simpler discussion restricting on the $\nu$SM part only. Though the present paper does not address the UV-extended DMP~\cite{Minakata:2021nii}, the treatment of the UV part is useful if one wants to extend our present discussion to the theory with non-unitarity~\cite{Antusch:2006vwa,Escrihuela:2015wra,Blennow:2016jkn,Fong:2016yyh,Fong:2017gke}. See e.g., refs.~\cite{Antusch:2006vwa,Escrihuela:2015wra,Blennow:2016jkn,Fong:2016yyh,Fong:2017gke,Minakata:2021nii,Minakata:2022zua}, and the references therein for an incomplete list of the papers on UV. 

We take the uniform matter density approximation in this paper. 
Though it may make our point of emphasizing the local nature of the Rep transformation less impressive, the locality is already obvious from the Hamiltonian proof of the symmetry~\cite{Minakata:2021dqh,Minakata:2022zua}. It is also not difficult to extend our discussion to the case of adiabatically varying matter density. 

\subsection{The flavor basis $S$ matrix: Non-perturbative form} 
\label{sec:flavor-S} 

Under the uniform matter density approximation the bar-basis $\bar{S}$ matrix is given by
\begin{eqnarray} 
\bar{S} (x) = e^{-i \bar{H} x}. 
\label{hat-Smatrix}
\end{eqnarray}
Using the property 
\begin{eqnarray} 
&&
U_{23} U_{13}(\phi) U_{12} (\psi) 
\bar{S} 
U_{12}^{\dagger} (\psi) U^\dagger_{13}(\phi) U_{23}^{\dagger} 
= 
e^{-i  U_{23} U_{13}(\phi) U_{12} (\psi) 
\bar{H} 
U_{12}^{\dagger} (\psi) U^\dagger_{13}(\phi) U_{23}^{\dagger}  x } 
\nonumber 
\end{eqnarray}
and noticing eq.~\eqref{H-S-flavor-bar}, the flavor-basis $S$ matrix reads
\begin{eqnarray} 
S_{ \text{flavor} } 
&=& 
\exp{ \left[ - i U_{23} U_{13}(\phi) U_{12} (\psi) 
\bar{H} 
U_{12}^{\dagger} (\psi) U^\dagger_{13}(\phi) U_{23}^{\dagger}  x \right] }, 
\label{S-flavor}
\end{eqnarray}
which is equal to $e^{ - i H_{ \text{flavor} } x}$, as it should. 

\subsection{Transformation property of the flavor basis $S$ matrix} 
\label{sec:transf-S} 

Noticing that the zeroth-order $V$ matrix is given by 
$V^{(0)} ( \theta_{23}, \phi, \psi, \delta) = U_{23} U_{13}(\phi) U_{12} (\psi, \delta)$,  
$H_{ \text{flavor} } $ can be written as 
\begin{eqnarray}
&&
H_{ \text{flavor} } 
= 
V^{(0)} ( \theta_{23}, \phi, \psi, \delta) 
\bar{H} ( \theta_{23}, \theta_{12}, \phi, \psi, \delta;  \lambda_{i} ) 
\left[ V^{(0)} ( \theta_{23}, \phi, \psi, \delta) \right]^{\dagger}. 
\label{H-flavor-by-V}
\end{eqnarray}
In the discussion that follows, the identity~\cite{Minakata:2022zua} 
\begin{eqnarray}
&&
V^{(0)} ( \theta_{23}^{\prime}, \phi^{\prime}, \psi^{\prime}, \delta^{\prime}) 
R 
\left[ V^{(0)} ( \theta_{23}, \phi, \psi, \delta) \right]^{\dagger} 
= \text{Rep(X)} 
\label{identity}
\end{eqnarray}
is of key importance, where the rephasing factor Rep(X) is given by 
Rep(I) = diag (1,1,1), and 
\begin{eqnarray} 
&&
\text{Rep(II)} = 
\left[
\begin{array}{ccc}
1 & 0 & 0 \\
0 & -1 & 0 \\
0 & 0 & 1 \\
\end{array}
\right], 
\hspace{6mm}
\text{Rep(III)} = 
\left[
\begin{array}{ccc}
- 1 & 0 & 0 \\
0 & 1 & 0 \\
0 & 0 & 1
\end{array}
\right], 
\hspace{6mm}
\text{Rep(IV)} = 
\left[
\begin{array}{ccc}
- 1 & 0 & 0 \\
0 & -1 & 0 \\
0 & 0 & 1
\end{array}
\right]. 
\label{Rep-II-III-IV} 
\end{eqnarray} 
Using the identity one can write down the transformation property of $H_{ \text{flavor} }$ under Symmetry X as 
\begin{eqnarray}
&&
H_{ \text{flavor} } 
\rightarrow_{\text{\tiny Symmetry X}} 
V^{(0)} ( \theta_{23}^{\prime}, \phi^{\prime}, \psi^{\prime}, \delta^{\prime} ) 
\bar{H} ( \theta_{23}^{\prime}, \theta_{12}^{\prime}, \phi^{\prime}, \psi^{\prime}, \delta^{\prime}; \lambda_{i}^{\prime} ) 
\left[ V^{(0)} ( \theta_{23}^{\prime}, \phi^{\prime}, \psi^{\prime}, \delta^{\prime} ) \right]^{\dagger} 
\nonumber \\
&& 
\hspace{-8mm} 
=  
\text{Rep(X)} 
V^{(0)} ( \theta_{23}, \phi, \psi, \delta) R^{\dagger} 
\bar{H} ( \theta_{23}^{\prime}, \theta_{12}^{\prime}, \phi^{\prime}, \psi^{\prime}, \delta^{\prime}; \lambda_{i}^{\prime} ) 
R 
\left[ V^{(0)} ( \theta_{23}, \phi, \psi, \delta) \right]^{\dagger} 
\text{Rep(X)} ^{\dagger}.
\label{H-transf} 
\end{eqnarray}
Since it is shown in ref.~\cite{Minakata:2022zua} that 
\begin{eqnarray} 
&& 
R^{\dagger} 
\bar{H} ( \theta_{23}^{\prime}, \theta_{12}^{\prime}, \phi^{\prime}, \psi^{\prime}, \delta^{\prime}; \lambda_{i}^{\prime} ) 
R 
= 
\bar{H} ( \theta_{23}, \theta_{12}, \phi, \psi, \delta; \lambda_{i} )
\label{RHR}
\end{eqnarray} 
for all Symmetry X-DMP-UV where X= IA, IB, $\cdot \cdot \cdot $, IVB, the transformation property of $H_{ \text{flavor} } $ under Symmetry X reads: 
\begin{eqnarray}
&&
H_{ \text{flavor} } \rightarrow_{\text{\tiny Symmetry X}} 
\text{Rep(X)} H_{ \text{flavor} } 
\text{Rep(X)} ^{\dagger}. 
\label{H-flavor-transf}
\end{eqnarray}
It implies that the transformation property of $S_{ \text{flavor} }$ under Symmetry X is given by 
\begin{eqnarray}
S_{ \text{flavor} } 
\rightarrow
\text{ Rep(X)} S_{ \text{flavor} } \text{ Rep(X)}^{\dagger}. 
\label{S-flavor-transf}
\end{eqnarray}

If we parametrize $S_{ \text{flavor} }$ as
\begin{eqnarray} 
&&
S_{ \text{flavor} } 
=
\left[
\begin{array}{ccc}
S_{e e} & S_{e \mu} & S_{e \tau} \\
S_{\mu e} & S_{\mu \mu} & S_{\mu \tau} \\
S_{\tau e} & S_{\tau \mu} & S_{\tau \tau} \\
\end{array}
\right], 
\label{S-flavor-param}
\end{eqnarray}
some of the $S_{ \text{flavor} }$ matrix elements $S_{\alpha \beta}$ flip sign under Symmetry X, according to eq.~\eqref{S-flavor-transf}: 
\begin{eqnarray}
&&
\text{X = II:}  
\hspace{10mm} 
S_{e \mu} \rightarrow - S_{e \mu} 
\hspace{10mm} 
S_{\tau \mu} \rightarrow - S_{\tau \mu}, 
\nonumber \\
&&
\text{X = III:}  
\hspace{10mm} 
S_{e \mu} \rightarrow - S_{e \mu} 
\hspace{10mm} 
S_{\tau e} \rightarrow - S_{\tau e}, 
\nonumber \\
&&
\text{X = IV:}  
\hspace{10mm} 
S_{e \tau} \rightarrow - S_{e \tau} 
\hspace{10mm} 
S_{\tau \mu} \rightarrow - S_{\tau \mu}, 
\label{Sab-flip-sign}
\end{eqnarray}
and no sign change for X=I as Rep(I)=1. 
The result is perfectly consistent with the sign flip for X=II and III in $S_{e \mu}$, and the sign flip for X=II and IV in $S_{\tau \mu}$, as observed in section~\ref{sec:analytic-mapping}. Notice that the phase factors $e^{ i (\Lambda_{j}/2E) x}$ ($j=1,2$) to define $S_{\alpha \beta}^{ \text{Reph-1} }$ and $S_{\alpha \beta}^{ \text{Reph-2} }$ do not affect the sign flip. 

\subsection{A perturbative study of eigenvalue exchange $\Lambda_{1} \leftrightarrow \Lambda_{2}$}
\label{sec:Lambda-exchange}

To make progress in the step (2b), we calculate the higher-order corrections of the energy eigenvalues using the steady-state perturbation theory~\cite{Landau:1981} with $\bar{H}$ in eq.~\eqref{barH-0th-1st}. Thanks to the special structure of the Hamiltonian, the first- and the third-order corrections vanish, the feature which prevails to all odd-order corrections~\cite{Denton:2019ovn}. To fourth order in expansion, the results can be expressed by using the compact notation $\widetilde{\epsilon} \equiv \epsilon c_{12} s_{12} s_{( \phi - \theta_{13} )}$ as
\begin{eqnarray} 
&&
\Lambda_{1} 
= 
\lambda_{1} 
- 
\left( \widetilde{\epsilon} \Delta m^2_{ \text{ren} } \right)^2   
\frac{ s_\psi^2 }{ \lambda_{3} - \lambda_{1} } 
+ 
\left( \widetilde{\epsilon} \Delta m^2_{ \text{ren} } \right)^4 
\biggl\{ 
\frac{ s^4_\psi }{ ( \lambda_{3} - \lambda_{1} )^3 } 
- \frac{ c^2_\psi s^2_\psi }{ ( \lambda_{3} - \lambda_{1} )^2 ( \lambda_{2} - \lambda_{1} ) } 
\biggr\}, 
\nonumber \\
&&
\Lambda_{2} 
= 
\lambda_{2} 
- \left( \widetilde{\epsilon} \Delta m^2_{ \text{ren} } \right)^2  
\frac{ c_\psi^2 }{ \lambda_{3} - \lambda_{2} } 
+ 
\left( \widetilde{\epsilon} \Delta m^2_{ \text{ren} } \right)^4 
\biggl\{ 
\frac{ c^4_\psi }{ ( \lambda_{3} - \lambda_{2} )^3 } 
+ \frac{ c^2_\psi s^2_\psi }{ ( \lambda_{3} - \lambda_{2} )^2 ( \lambda_{2} - \lambda_{1} ) } 
\biggr\}, 
\nonumber \\
&&
\Lambda_{3} 
= 
\lambda_{3} 
+ 
\left( \widetilde{\epsilon} \Delta m^2_{ \text{ren} } \right)^2  
\biggl\{ \frac{ s_\psi^2 } { \lambda_{3} - \lambda_{1} } 
+ \frac{ c_\psi^2 } { \lambda_{3} - \lambda_{2} } \biggr\} 
\nonumber \\
&-& 
\left( \widetilde{\epsilon} \Delta m^2_{ \text{ren} } \right)^4 
\biggl\{ 
\frac{ s^4_\psi }{ ( \lambda_{3} - \lambda_{1} )^3 } 
+ \frac{ c^4_\psi }{ ( \lambda_{3} - \lambda_{2} )^3 } 
+ \frac{ c^2_\psi s^2_\psi  }{ ( \lambda_{3} - \lambda_{1} )^2 ( \lambda_{3} - \lambda_{2} ) } 
+ \frac{ c^2_\psi s^2_\psi  }{ ( \lambda_{3} - \lambda_{1} ) ( \lambda_{3} - \lambda_{2} )^2 } 
\biggr\}.
\nonumber \\
\label{Lambda-4th}
\end{eqnarray} 
The second-order corrections are computed in ref.~\cite{Denton:2016wmg}, and the reference~\cite{Denton:2019ovn} presents the fourth and sixth order corrections to $\Lambda_{1}$. It is suggested to obtain $\Lambda_{2}$ by using Symmetry IA-DMP, a valid procedure, but it does not perfectly fit to our purpose of (2b). 

Using eq.~\eqref{Lambda-4th} it is easy to show that to the fourth-order expansion the Rep symmetry transformations in Table~\ref{tab:DMP-symmetry} induce the eigenvalue exchange $\Lambda_{1} \leftrightarrow \Lambda_{2}$ for all the Symmetry X-DMP (X=IA, IB, $\cdot \cdot \cdot $, IVB). $\Lambda_{3}$ stays untransformed. Unfortunately, demonstration of this property to all orders in the DMP perturbation theory is not with us at this moment: The task (2b) remains uncompleted.  

\section{Consistency of the perturbation theory of neutrino oscillation} 
\label{sec:consistency} 

The atmospheric-resonance perturbation theory developed in refs.~\cite{Arafune:1996bt,Cervera:2000kp,Freund:2001pn,Akhmedov:2004ny,Minakata:2015gra} treat the resonance region as isolated from the solar one. Then, an immediate question would arise in the light of our statement of the non-isolated resonance in sections~\ref{sec:one-res-one-sym} and~\ref{sec:Rep-what-is}. Does this feature imply any inconsistency of the perturbative framework? In this section, we present our heuristic discussion on convergence of the perturbative series in the globally- and locally-valid theories. It strongly suggests that all these theories have the common sound property as the perturbation theory, clearing up the above doubt at the level of robustness of our argument. 

Convergence of the perturbative series is an important but highly nontrivial subject in quantum mechanics and quantum field theories~\cite{Itzykson:1980rh}. Our treatment in this section, albeit with  limited scope using the energy eigenvalue sum rule, may be useful in more extended perspective. 

\subsection{Eigenvalue sum rule} 
\label{sec:sum-rule} 

If we denote the solutions of $\det \left[ 2EH - \Lambda \right] = 0$ as $\Lambda_{k}$ ($k=1,2,3$), it is well known that they must satisfy the three sum rules (characteristic equations), see e.g., refs.~\cite{Minakata:2015gra,Denton:2016wmg}: 
\begin{eqnarray} 
\Lambda_{1} + \Lambda_{2} + \Lambda_{3} 
&=& a + \Delta m^2_{31} + \Delta m^2_{21}, 
\nonumber \\
\Lambda_{1} \Lambda_{2} + \Lambda_{2} \Lambda_{3} + \Lambda_{3} \Lambda_{1} 
&=&
\Delta m^2_{21} \Delta m^2_{31} 
+ a \left\{ ( c^2_{12} + s^2_{12} s^2_{13} ) \Delta m^2_{21} + c^2_{13} \Delta m^2_{31} 
\right\}, 
\nonumber \\
\Lambda_{1} \Lambda_{2} \Lambda_{3}
&=& 
c^2_{12} c^2_{13} 
a \Delta m^2_{21} \Delta m^2_{31}. 
\label{sum-rule-exact}
\end{eqnarray} 
We will show below that these sum rules are obeyed {\em exactly} to fourth order in perturbation theory in the three perturbative frameworks, DMP~\cite{Denton:2016wmg}, SRP~\cite{Martinez-Soler:2019nhb}, and the helio-perturbation theory~\cite{Minakata:2015gra}. We then argue that it is natural to expect the property holds to all orders in perturbation theory. 

To give an idea to the readers, we first give a very rough sketch of the common features of our computation of the left-hand side of eq.~\eqref{sum-rule-exact} in these three perturbative frameworks. When the zeroth-order eigenvalues are used, the results do not agree with the exact ones in eq.~\eqref{sum-rule-exact}. But, the mismatch is of second order only, and it is cancelled out exactly when the second-order corrections in $\Lambda_{k}$ are taken into account, leaving the exact sum rule eq.~\eqref{sum-rule-exact}. It can be readily shown that the fourth-order correction terms cancel to each other, leaving the exact sum rule unaffected. 

\subsection{Sum rule in DMP and the other two theories} 
\label{sec:sum-rule-DMP} 

We show the explicit calculations in the DMP perturbation theory to some details. The features in SRP and the helio-perturbation theory exactly mimic those of DMP. If we use the zeroth-order eigenvalues $\lambda_{k}$, the sum rule eq.~\eqref{sum-rule-exact} becomes 
\begin{eqnarray} 
\lambda_{1} + \lambda_{2} + \lambda_{3} 
&=& 
a + \Delta m^2_{31} + \Delta m^2_{21}, 
\nonumber \\
\lambda_{1} \lambda_{2} + \lambda_{2} \lambda_{3} + \lambda_{3} \lambda_{1} 
&=&
\Delta m^2_{21} \Delta m^2_{31} 
+ a \left\{ ( c^2_{12} + s^2_{12} s^2_{13} ) \Delta m^2_{21} + c^2_{13} \Delta m^2_{31} 
\right\} 
\nonumber \\
&+& 
c^2_{12} s^2_{12} s^2_{( \phi - \theta_{13} )} 
\left( \epsilon \Delta m^2_{ \text{ren} } \right)^2, 
\nonumber \\
\lambda_{1} \lambda_{2} \lambda_{3}
&=& 
c^2_{12} ( \epsilon \Delta m^2_{ \text{ren} } ) 
\biggl\{ 
s^2_{12} ( \epsilon \Delta m^2_{ \text{ren} } ) 
\left( \Delta m^2_{ \text{ren} } + a \right) 
+ c^2_{13} a \Delta m^2_{ \text{ren} } 
\biggr\} 
\nonumber \\
&-& 
\frac{ 1 }{ 2 } 
c^2_{12} s^2_{12} c^2_{( \phi - \theta_{13} )} 
\left( \epsilon \Delta m^2_{ \text{ren} } \right)^2 
\biggl\{ 
\left( \Delta m^2_{ \text{ren} } + a \right) + \sqrt{ \left( \Delta m^2_{ \text{ren} } - a \right)^2 + 4 s^2_{13} a \Delta m^2_{ \text{ren} } } 
\biggr\} 
\nonumber \\
&+& 
c^2_{12} s^4_{12} s^2_{( \phi - \theta_{13} )} \left( \epsilon \Delta m^2_{ \text{ren} } \right)^3.  
\label{sum-rule-DMP}
\end{eqnarray} 
As we see the first sum rule is exact and one can show that no second and higher order correction is induced in all the three theories, DMP, SRP, and the helio-perturbation theories. Therefore, we discuss the second and third sum rules in our discussion below, in which we have only the second-order correction terms, apart from an exceptional third-order one in the third sum rule. Of course, existence of the correction terms itself is normal as these frameworks are not meant to be exact. 

We next discuss the sum rule with adding only the second-order corrections to the eigenvalues. First of all, the order $( \epsilon \Delta m^2_{ \text{ren} } )^3$ term in the last line in eq.~\eqref{sum-rule-DMP} has to be absent, because the odd-order corrections to the eigenvalues should vanish~\cite{Denton:2019ovn}. A computation shows that $\Lambda_{1} \Lambda_{2} \Lambda_{3} = \lambda_{1} \lambda_{2} \lambda_{3} - c^2_{12} s^2_{12} s^2_{( \phi - \theta_{13} )} \left( \epsilon \Delta m^2_{ \text{ren} } \right)^2 \lambda_{-}$, where $\lambda_{-}$\footnote{
%%%%%%%%%%%% footnote %%%%%%%%%%%%%
$\lambda_{-} = \frac{ 1 }{ 2 } \biggl\{ \left( \Delta m^2_{ \text{ren} } + a \right) - \sqrt{ \left( \Delta m^2_{ \text{ren} } - a \right)^2 + 4 s^2_{13} a \Delta m^2_{ \text{ren} } } \biggr\} + s^2_{12} (\epsilon \Delta m^2_{ \text{ren} })$. }
denotes one of the eigenvalues which participates the atmospheric level crossing, see refs.~\cite{Minakata:2015gra,Denton:2016wmg}. The order $( \epsilon \Delta m^2_{ \text{ren} } )$ component in $\lambda_{-}$ produces an $( \epsilon \Delta m^2_{ \text{ren} } )^3$ term which just cancels out the $( \epsilon \Delta m^2_{ \text{ren} } )^3$ term from $\lambda_{1} \lambda_{2} \lambda_{3}$.  
If we plug in the expressions of $\Lambda_{k}$ to second-order corrections into the second sum rule, we obtain  
\begin{eqnarray} 
&&  
\Lambda_{1} \Lambda_{2} + \Lambda_{2} \Lambda_{3} +\Lambda_{3} \Lambda_{1} 
= 
\lambda_{1} \lambda_{2} + \lambda_{2} \lambda_{3} + \lambda_{3} \lambda_{1} 
- c^2_{12} s^2_{12} s^2_{( \phi - \theta_{13} )} 
\left( \epsilon \Delta m^2_{ \text{ren} } \right)^2, 
\label{2nd-sum-rule}
\end{eqnarray} 
whose correction term exactly cancel the one in the DMP sum rule in eq.~\eqref{sum-rule-DMP}. Then, the exact second sum rule in eq.~\eqref{sum-rule-exact} is reproduced. The cancellation in the third sum rule is not that simple, but it is not difficult to show that all the collections terms cancel out, leaving the exact third sum rule in eq.~\eqref{sum-rule-exact}. 

We should remark here that if the all-order DMP respects the eigenvalue sum rule, it has to be second-order exact. It must be the case because the fourth-order terms cannot cancel the second-order deviation from the exact sum rule.\footnote{
%%%%%%%%%%%% footnote %%%%%%%%%%%%
This important point is remarked to the author by Peter Denton to whom he gratefully acknowledge. }
Therefore, when the fourth-order corrections to the eigenvalues are included, all the terms proportional to $( \epsilon \Delta m^2_{ \text{ren} } )^4$, including such terms as induced from the second order terms of $\Lambda_{k}$, do nothing but to cancel with each other to vanish. This feature is indeed confirmed by the explicit calculations. 

The eigenvalues to the fourth-order in perturbation theory in SRP~\cite{Martinez-Soler:2019nhb} and the helio-perturbation~\cite{Minakata:2015gra} theories are provided in Appendix~\ref{sec:eigenvalues-to-4thorder}. Using these eigenvalues, it is not difficult to show that the features that we have just observed in DMP, the second-order exact eigenvalue sum rule and vanishing fourth-order corrections, prevail in SRP and the helio-perturbation theory. 
Since this feature, which is common in all the three theories, is so charming we tend to speculate that the perturbative series converges to yield the exact energy eigenvalue sum rule in eq.~\eqref{sum-rule-exact} in DMP, SRP, and the helio-perturbation theories. Therefore, we have found no indication of the above mentioned suspicion, or unsound feature, in the helio-perturbation theory, and in the other two theories. 

Though the above-given argument itself cannot be regarded as an all-order proof of convergence of the perturbation series and the exact sum rule~\eqref{sum-rule-exact},\footnote{
%%%%%%%%%%%% footnote %%%%%%%%%%%%% 
An all-order proof would necessitates to show that there is no higher order corrections to the eigenvalue sum rule. In this respect it would be interesting to demonstrate the cancellation between $( \epsilon \Delta m^2_{ \text{ren} } )^{2n}$ terms in the sum rules to workable higher orders of $n \geq 3$. }
there exists a circumstantial support for our argument of the convergence in DMP. A numerical examination of the convergence of the DMP perturbation series is carried out by the authors of ref.~\cite{Denton:2019ovn}. A more recent study to the twelveth order shows, very roughly speaking, a factor of $\sim10^{6}$ improvement in the relative accuracy of the eigenvalue in each step of $m$-th to $(m+2)$-th orders~\cite{Parke:2023}. 

Recently we have found that the ZS system in matter possesses the eight 1-2 state exchange symmetries, Symmetry X-ZS (X=IA, IB, $\cdot \cdot \cdot $, IVB)~\cite{Minakata:2022zua}. Then, an intriguing question would be whether the eight DMP Rep symmetries~\cite{Minakata:2021dqh} fuses into Symmetry X-ZS when the perturbation series is summed, assuming convergence, to all orders. 
This picture is argued to be realized in the two-flavor toy model in ref.~\cite{Minakata:2021goi}. 

\section{On quantum mechanical formulation of neutrino oscillation }
\label{sec:QM-QFT} 

It is worth to call the readers' attention to the fact that our quantum mechanical framework for describing neutrino oscillation is an extremely restricted one, as discussed in ref.~\cite{Minakata:2022zua}. That is, it uses the approximation that neutrinos undergo no inelastic scattering, no absorption, and they are stable particles. Then, the quantum state space becomes a direct product of the one that has definite neutrino energy $E$. The system has only the three component states in each energy subspace, the neutrino flavor states $[\nu_{e}, \nu_{\mu}, \nu_{\tau}]^T$, or the mass eigenstates $[\nu_{1}, \nu_{2}, \nu_{3}]^T$, which is analogous to the spin 1 system. The flavor mixing matrix $U$ multiplied to the three-component states is nothing but the matrix representation of the ``rotation'' operator that acts onto the quantum state, which can be represented by the Gell-Mann matrix~\cite{Itzykson:1980rh}. The Rep symmetry, an invariance under the relatively simple discrete transformations of the quantum variables, resides in such greatly reduced quantum system. 

In the other extreme, neutrinos may be fully interacting by themselves, undergoing elastic or inelastic scatterings, producing particles by weak interactions, or absorbed into the environment. Then, the above framework no longer holds. If the ultra-relativistic approximation for neutrinos does not apply, the right framework should respect the Lorentz (or Pioncar\'e) invariance in a manifest way. In these cases the system is to be described by relativistic quantum field theory (QFT). It is a tantalizing challenge to identify evidences for the QFT effect beyond the above, commonly taken quantum mechanical framework. It still eludes us at this moment to the author's knowledge. From theoretical point of view, it is discussed that the $S$ matrix formulation in QFT framework is useful in resolving the paradoxical issues in theory of neutrino oscillation. See refs.~\cite{Akhmedov:2009rb,Akhmedov:2010ua}, and the references cited therein. 

Due to vastly different feature of the QFT description of neutrino systems in more generic environments from the quantum mechanical one, it appears too hard to find the answer to the question of whether our Rep symmetry has an extension in such regime. A step by step approach seems needed toward answering the question. In ref.~\cite{Minakata:2021dqh} we have suggested that a mean-field treatment of high-density neutrino gases may bear a close resemblance to our background-matter treatment of the neutrino system. To gain insight on physics and symmetry in interacting neutrino systems, one may start to work on high-density neutrino gases, though the task is far beyond the scope of this paper.  

\section{Concluding remarks}
\label{sec:conclusion} 

It appears that the reparametrization (Rep) symmetry discussed in this paper exists (almost) universally in the various perturbative frameworks, and even in the exact ZS construction in matter. It strongly suggests that a certain general principle is behind the Rep symmetry. In this paper we have argued that it is the $S$ matrix rephasing invariance which is tied up with the probabilistic interpretation of quantum mechanics. In fact, the intimate relationship between the rephasing invariance and the Rep symmetry is very noticeable throughout this paper. Given the nature of the rephasing invariance it strongly suggests that the Rep symmetry is rooted deep in quantum mechanics. If this picture is firmly established, it should provide a clear answer to the question: ``For what reason does the Rep symmetry exist?''. 

In Introduction we have started our description of the $S$ matrix rephasing by emphasizing that it can take a non-local form (i.e., depend upon many spacial points), as well as a local one with the dynamical variables, the ones to be elevated to the operators in the bra-ket formulation of quantum mechanics~\cite{Dirac:1958}. The rephasing of the $S$ matrix is nontrivial even with the simple examples introduced in section~\ref{sec:S-rephased-emu}, $S_{\alpha \beta}^{ \text{Reph-1} } \equiv e^{ i (\lambda_{1}/2E) x} S_{\alpha \beta}$ and $S_{\alpha \beta}^{ \text{Reph-2} } \equiv e^{ i (\lambda_{2}/2E) x} S_{\alpha \beta}$, where $\lambda_{k}/2E$ denotes the k-th eigenvalue of the Hamiltonian. It would not be obvious whether the explicit expressions of $S_{\alpha \beta}^{ \text{Reph-1} }$ and $S_{\alpha \beta}^{ \text{Reph-2} }$ in eqs.~\eqref{S-emu-Reph1} and ~\eqref{S-emu-Reph2} (or eqs.~\eqref{S-taumu-Reph1-SM} and~\eqref{S-taumu-Reph2-SM}) produce the same observable, unless we know a priori that it must be enforced by the probabilistic interpretation. 
In our new technology with the Rep symmetry, however, the physical equivalence between $S_{\alpha \beta}^{ \text{Reph-1} }$ and $S_{\alpha \beta}^{ \text{Reph-2} }$ is evident: The symmetry transformations map $S_{\alpha \beta}^{ \text{Reph-1} }$ to $S_{\alpha \beta}^{ \text{Reph-2} }$, or vice versa. This is an entirely new way of demonstrating the $S$ matrix rephasing invariance to the best of our knowledge, which in turn illuminates the Rep symmetry's quantum mechanical nature. 

In fact, the reverse way of the argument is also possible. For a given Rep symmetry of the state-exchange type which leaves the $S$ matrix invariant up to a factor of $\pm 1$, one can construct the $\xi$ parameter extension of the two rephased amplitudes $S_{\alpha \beta}^{ \text{Reph-1} } (\xi)$ and $S_{\alpha \beta}^{ \text{Reph-2} } (\xi)$, see eq.~\eqref{Reph-12-family}, which can be mapped to each other. 
However, we must note that the two-ways discussions showing the intimate connection between the $S$ matrix rephasing invariance and the Rep symmetry is valid only to the first order in the DMP perturbation theory. Therefore, an all-order construction of the $\xi$ parameter extended treatment of the $S$ matrix rephasing invariance is attempted, in which this reverse way plays an important role. Nonetheless, the task remains uncompleted due to the difficulty in proving the exchange property of the all-order eigenvalues $\Lambda_{1}$ and $\Lambda_{2}$. To show the consistency of our argument, we have verified this property to the fourth order in the perturbation theory. 

The other noticeable feature in the intimate relationships between the $S$ matrix rephasing invariance and the Rep symmetry is that there is a case that the former cannot be represented locally by the latter, which is related to the apparent absence of 1-3 exchange symmetry in DMP, see section~\ref{sec:rephasing-wider}. But this problem, once interpreted correctly, seems to illuminate the interesting difference between the features of the solar-scale and atmospheric-scale resonances. While the solar resonance can be dealt with as an isolated resonance in a good approximation, the atmospheric resonance is influenced by the solar one and affected by its small but non-negligible perturbation. If this is the correct interpretation, it is remarkable that this feature is indeed ``detected'' by the Rep symmetry, by indicating the absence of the pure 1-3 (or 2-3) exchange symmetry in DMP. 

Finally, we have asked ourselves the question of whether the perturbative treatment of the atmospheric resonance as an isolated object would imply any inconsistency in the theory. We have argued by using the eigenvalue sum rule that the perturbative series converges and the exact sum rule holds in the globally- and locally-valid frameworks. To the level of robustness of our discussion, it sweeps away the above doubt. Thus, we have revealed a new usage of the Rep symmetry as a diagnosing tool of the neutrino oscillation theory. 

What is a possible implication for the quantum mechanical nature of the Rep symmetry? In ref.~\cite{Minakata:2022zua}, we have discussed the Rep symmetry in an extended setting of the DMP theory with non-unitary flavor mixing matrix, or UV (unitarity violation). Of course, quantum mechanics governs the both $\nu$SM and UV parts of the theory. Then, it is natural to expect that the quantum-mechanics-rooted Rep symmetry transformations connect the dynamical variables in the both $\nu$SM and UV parts. In this way a low-energy description of the relationship between the $\nu$SM and UV sectors may become possible, as attempted in ref.~\cite{Minakata:2022zua}. A picture of the inter-sector communication through CP phases has emerged.

\begin{acknowledgments} 
Our treatment of the energy sum rule is inspired by Peter Denton to whom the author would like to acknowledge. He thanks Stephen Parke for informative correspondences on convergence of the perturbation series. 

\end{acknowledgments}

\appendix 

\section{$S$ matrices in the $\nu_{\mu} \rightarrow \nu_{\tau}$ channel}
\label{sec:S-tau-mu} 

The $S$ matrix element in the $\nu_{\mu} \rightarrow \nu_{\tau}$ channel to the first order in the DMP perturbation theory reads: 
\begin{eqnarray} 
&& 
S_{\tau \mu} 
= 
%\nonumber \\ &=& 
- \left( c^2_{23} e^{ i \delta} - s^2_{23} e^{ - i \delta} \right) 
s_{\phi} c_\psi s_\psi \left( e^{ - i h_{2} x }  - e^{ - i h_{1} x } \right) 
\nonumber \\ 
&+&
c_{23} s_{23} 
\left[ 
( s^2_{\phi} s^2_\psi - c^2_\psi ) e^{ - i h_{2} x } 
+ ( s^2_{\phi} c^2_\psi - s^2_\psi ) e^{ - i h_{1} x } 
+ c^2_{\phi} e^{ - i h_{3} x } 
\right] 
\nonumber \\ 
&+& 
\epsilon c_{12} s_{12} s_{ (\phi - \theta_{13}) } 
\left( c^2_{23} e^{ i \delta} - s^2_{23} e^{ - i \delta} \right) c_{\phi} 
\left[ 
c^2_{\psi} 
\frac{ \Delta_{ \text{ren} } }{ h_{3} - h_{2} } 
\left( e^{ - i h_{3} x } - e^{ - i h_{2} x } \right) 
+ s^2_{\psi} 
\frac{ \Delta_{ \text{ren} } }{ h_{3} - h_{1} } 
\left( e^{ - i h_{3} x } - e^{ - i h_{1} x } \right)  
\right] 
\nonumber \\
&-&
2 \epsilon c_{12} s_{12} s_{ (\phi - \theta_{13}) } 
c_{23} s_{23} c_{\phi} s_{\phi} c_{\psi} s_\psi 
\left[ 
\frac{ \Delta_{ \text{ren} } }{ h_{3} - h_{2} } 
\left( e^{ - i h_{3} x } - e^{ - i h_{2} x } \right) 
- \frac{ \Delta_{ \text{ren} } }{ h_{3} - h_{1} } 
\left( e^{ - i h_{3} x } - e^{ - i h_{1} x } \right) 
\right]. 
\label{S-taumu-SM} 
\end{eqnarray}
Using the definitions in eq.~\eqref{S-Reph-12-def}, the expressions of $S_{\tau \mu}^{ \text{Reph-1} }$ and $S_{\tau \mu}^{ \text{Reph-2} }$ are given by 
\begin{eqnarray} 
&& 
S_{\tau \mu}^{ \text{Reph-1} } 
\nonumber \\
&=& 
c_{23} s_{23} c^2_{\phi} \left\{ e^{ - i ( h_{3} - h_{1} ) x } - 1 \right\} 
- \left[ 
\left( c^2_{23} e^{ i \delta} - s^2_{23} e^{ - i \delta} \right) s_{\phi} c_\psi s_\psi 
- c_{23} s_{23} ( s^2_{\phi} s^2_\psi - c^2_\psi ) 
\right] 
\left\{ e^{ - i ( h_{2} - h_{1} ) x } - 1 \right\} 
\nonumber \\
&-& 
\epsilon c_{12} s_{12} s_{ (\phi - \theta_{13}) } 
\left[ 
\left( c^2_{23} e^{ i \delta} - s^2_{23} e^{ - i \delta} \right) c_{\phi} c^2_{\psi} 
- 2 c_{23} s_{23} c_{\phi} s_{\phi} c_{\psi} s_\psi 
\right] 
\frac{ \Delta_{ \text{ren} } }{ h_{3} - h_{2} } 
\left\{ e^{ - i ( h_{2} - h_{1} ) x } - 1 \right\} 
\nonumber \\ 
&+& 
\epsilon c_{12} s_{12} s_{ (\phi - \theta_{13}) } 
\left[ 
\left( c^2_{23} e^{ i \delta} - s^2_{23} e^{ - i \delta} \right) c_{\phi} c^2_{\psi} 
- 2 c_{23} s_{23} c_{\phi} s_{\phi} c_{\psi} s_\psi 
\right] 
\frac{ \Delta_{ \text{ren} } }{ h_{3} - h_{2} } 
\left\{ e^{ - i ( h_{3} - h_{1} ) x } - 1 \right\} 
\nonumber \\
&+& 
\epsilon c_{12} s_{12} s_{ (\phi - \theta_{13}) } 
\left[ 
\left( c^2_{23} e^{ i \delta} - s^2_{23} e^{ - i \delta} \right) c_{\phi} 
s^2_{\psi} 
+ 2 c_{23} s_{23} c_{\phi} s_{\phi} c_{\psi} s_\psi 
\right] 
\frac{ \Delta_{ \text{ren} } }{ h_{3} - h_{1} } 
\left\{ e^{ - i ( h_{3} - h_{1} ) x } - 1 \right\}. 
\label{S-taumu-Reph1-SM} 
\end{eqnarray} 
\begin{eqnarray} 
&& 
S_{\tau \mu}^{ \text{Reph-2} } 
\nonumber \\
&=& 
c_{23} s_{23} c^2_{\phi} \left\{ e^{ - i ( h_{3} - h_{2} ) x } - 1 \right\} 
+ \left[ 
\left( c^2_{23} e^{ i \delta} - s^2_{23} e^{ - i \delta} \right) s_{\phi} c_\psi s_\psi 
+ c_{23} s_{23} ( s^2_{\phi} c^2_\psi - s^2_\psi ) 
\right] 
\left\{ e^{ i ( h_{2} - h_{1} )  x } - 1 \right\} 
\nonumber \\ 
&-& 
\epsilon c_{12} s_{12} s_{ (\phi - \theta_{13}) } 
\biggl[ 
\left( c^2_{23} e^{ i \delta} - s^2_{23} e^{ - i \delta} \right) c_{\phi} s^2_{\psi} 
+ 2 c_{23} s_{23} c_{\phi} s_{\phi} c_{\psi} s_\psi 
\biggr] 
\frac{  \Delta_{ \text{ren} } }{ h_{3} - h_{1} } 
\left\{ e^{ i ( h_{2} - h_{1} ) x } - 1 \right\} 
\nonumber \\ 
&+& 
\epsilon c_{12} s_{12} s_{ (\phi - \theta_{13}) } 
\left[ 
\left( c^2_{23} e^{ i \delta} - s^2_{23} e^{ - i \delta} \right) c_{\phi} 
c^2_{\psi}  
- 2 c_{23} s_{23} c_{\phi} s_{\phi} c_{\psi} s_\psi 
\right] 
\frac{ \Delta_{ \text{ren} } }{ h_{3} - h_{2} } 
\left\{ e^{ - i ( h_{3} - h_{2} ) x } - 1 \right\} 
\nonumber \\ 
&+& 
\epsilon c_{12} s_{12} s_{ (\phi - \theta_{13}) } 
\biggl[ 
\left( c^2_{23} e^{ i \delta} - s^2_{23} e^{ - i \delta} \right) c_{\phi} s^2_{\psi} 
+ 2 c_{23} s_{23} c_{\phi} s_{\phi} c_{\psi} s_\psi 
\biggr] 
\frac{  \Delta_{ \text{ren} } }{ h_{3} - h_{1} } 
\left\{ e^{ - i ( h_{3} - h_{2} ) x } - 1 \right\}. 
\label{S-taumu-Reph2-SM} 
\end{eqnarray}

Following the same procedure as in the $\nu_{\mu} \rightarrow \nu_{e}$ channel, it is easy to derive the result 
$S_{\alpha \beta}^{ \text{Reph-2} } \rightarrow \pm S_{\alpha \beta}^{ \text{Reph-1} }$ 
described in eq.~\eqref{mapping-emu} for $(\alpha \beta) = (\tau \mu)$, and the $\pm$ signs described in the text. 

\section{The eigenvalues to fourth-order corrections in SRP and the helio-perturbation theories} 
\label{sec:eigenvalues-to-4thorder}

The eigenvalues can be computed similarly in SRP~\cite{Martinez-Soler:2019nhb} with the results 
\begin{eqnarray} 
&& 
\Lambda_{1}
= 
\lambda_{1} 
- c^2_{13} s^2_{13} a^2 \frac{ c^2_{\varphi} }{ ( \lambda_{3} - \lambda_{1} ) } 
+ 
c^4_{13} s^4_{13} a^4 
\biggl\{ 
\frac{ c^4_{\varphi} }{ ( \lambda_{3} - \lambda_1 )^3 } 
- \frac{ c^2_{\varphi} s^2_{\varphi} }{ ( \lambda_{3} - \lambda_1 )^2 ( \lambda_{2} - \lambda_{1} ) } 
\biggr\} 
\nonumber \\
&& 
\Lambda_{2} 
= 
\lambda_{2} 
- c^2_{13} s^2_{13} a^2 \frac{ s^2_{\varphi} }{ ( \lambda_{3} - \lambda_{2} ) } 
+ 
c^4_{13} s^4_{13} a^4 
\biggl\{ 
\frac{ s^4_{\varphi} }{ ( \lambda_{3} - \lambda_{2} )^3 } 
+ \frac{ c^2_{\varphi} s^2_{\varphi} }{ ( \lambda_{3} - \lambda_{2} )^2 ( \lambda_{2} - \lambda_{1} ) } 
\biggr\} 
\nonumber \\
&& 
\Lambda_{3} 
= 
\lambda_{3} 
+ c^2_{13} s^2_{13} a^2 
\biggl\{ \frac{ c^2_{\varphi} }{ ( \lambda_{3} - \lambda_{1} ) } 
+ \frac{ s^2_{\varphi} }{ ( \lambda_{3} - \lambda_{2} ) } 
\biggr\} 
\nonumber \\
&-& 
c^4_{13} s^4_{13} a^4 
\biggl\{ 
\frac{ c^4_{\varphi} }{ ( \lambda_{3} - \lambda_1 )^3 } 
+ \frac{ s^4_{\varphi}  }{ ( \lambda_{3} - \lambda_{2} )^3 } 
+ \frac{ c^2_{\varphi} s^2_{\varphi} }{ ( \lambda_{3} - \lambda_1 ) ( \lambda_{3} - \lambda_{2} )^2 } 
+ \frac{ c^2_{\varphi} s^2_{\varphi} }{ ( \lambda_{3} - \lambda_{2} ) ( \lambda_{3} - \lambda_1 )^2 } 
\biggr\}, 
\nonumber \\
\label{eigenvalues-SRP}
\end{eqnarray} 
where the explicit expressions of SRP leading order eigenvalues $\lambda_{k}$ ($k=1,2,3$) and the definition of $\varphi$, the matter-affected $\theta_{12}$, follows the ones in ref.~\cite{Martinez-Soler:2019nhb}. 

Similarly, the eigenvalues to fourth-order corrections in the helio-perturbation theory~\cite{Minakata:2015gra} are given by 
\begin{eqnarray} 
\Lambda_{-} 
&=&
\lambda_{-} 
+ \frac{ c^2_{ (\phi - \theta_{13}) } }{ ( \lambda_{-} - \lambda_0 ) } 
c^2_{12} s^2_{12} ( \epsilon \Delta m^2_{ \text{ren} } )^2 
+ \biggl\{ 
- \frac{ c^4_{ ( \phi - \theta_{13} ) } }{ ( \lambda_{-} - \lambda_{0} )^3 } 
+\frac{ c^2_{ ( \phi - \theta_{13} ) } s^2_{ ( \phi - \theta_{13} ) } }{ ( \lambda_{-} - \lambda_{0} )^2 ( \lambda_{-} - \lambda_{+} ) } 
\biggr\} 
c^4_{12} s^4_{12} ( \epsilon \Delta m^2_{ \text{ren} } )^4 
\nonumber \\
\Lambda_{+} 
&=&
\lambda_{+} 
+ \frac{ s^2_{ (\phi - \theta_{13}) } }{ ( \lambda_{+} - \lambda_0 ) } 
c^2_{12} s^2_{12} ( \epsilon \Delta m^2_{ \text{ren} } )^2 
+ 
\biggl\{ 
- \frac{ s^4_{ ( \phi - \theta_{13} ) } }{ ( \lambda_{+} - \lambda_{0} )^3 } 
+ \frac{ c^2_{ ( \phi - \theta_{13} ) } s^2_{ ( \phi - \theta_{13} ) }  }
{ ( \lambda_{+} - \lambda_{0} )^2 ( \lambda_{+} - \lambda_{-} ) } 
\biggr\} 
c^4_{12} s^4_{12} 
( \epsilon \Delta m^2_{ \text{ren} } )^4 
\nonumber \\
\Lambda_{0} 
&=&
\lambda_{0} 
- \frac{ c^2_{ (\phi - \theta_{13}) } }{ ( \lambda_{-} - \lambda_0 ) } 
c^2_{12} s^2_{12} ( \epsilon \Delta m^2_{ \text{ren} } )^2 
- \frac{ s^2_{ (\phi - \theta_{13}) } }{ ( \lambda_{+} - \lambda_0 ) } 
c^2_{12} s^2_{12} ( \epsilon \Delta m^2_{ \text{ren} } )^2 
\nonumber \\
&+&
\biggl\{ 
\frac{ c^4_{ ( \phi - \theta_{13} ) } }{ ( \lambda_{-} - \lambda_{0} )^3 } 
+ \frac{ s^4_{ ( \phi - \theta_{13} ) } }{ ( \lambda_{+} - \lambda_{0} )^3 } 
+ \frac{ c^2_{ ( \phi - \theta_{13} ) } s^2_{ ( \phi - \theta_{13} ) } }
{ ( \lambda_{-} - \lambda_{0} ) ( \lambda_{+} - \lambda_{0} )^2 } 
+ \frac{ c^2_{ ( \phi - \theta_{13} ) } s^2_{ ( \phi - \theta_{13} ) } }
{ ( \lambda_{-} - \lambda_{0} )^2 ( \lambda_{+} - \lambda_{0} ) } 
\biggr\} 
c^4_{12} s^4_{12} 
( \epsilon \Delta m^2_{ \text{ren} } )^4
\nonumber \\
\label{eigenvalues-helio-P}
\end{eqnarray} 
where the notations follow those of the original reference, ref.~\cite{Minakata:2015gra}. 
In the NMO, $\lambda_{-}$, $\lambda_{0}$, and $\lambda_{+}$ correspond respectively to $\lambda_{1}$, $\lambda_{2}$, and $\lambda_{3}$.


\begin{thebibliography}{99} 

%\cite{Dirac:1958}
\bibitem{Dirac:1958}
P.~A.~M.~Dirac, ``The Principles of Quantum Mechanics,'' Fourth Edition, 
Oxford University Press, 1958, 
ISBN 978-0198520115

%\cite{Wolfenstein:1977ue}
\bibitem{Wolfenstein:1977ue}
L.~Wolfenstein,
``Neutrino Oscillations in Matter,''
Phys. Rev. D \textbf{17} (1978), 2369-2374
doi:10.1103/PhysRevD.17.2369

%\cite{Mikheyev:1985zog} 
\bibitem{Mikheyev:1985zog}
S.~P.~Mikheyev and A.~Y.~Smirnov,
``Resonance Amplification of Oscillations in Matter and Spectroscopy of Solar Neutrinos,''
Sov. J. Nucl. Phys. \textbf{42} (1985), 913-917
 %%CITATION = SJNCA,42,913;%%

%\cite{Barger:1980tf} 
\bibitem{Barger:1980tf}
V.~D.~Barger, K.~Whisnant, S.~Pakvasa and R.~J.~N.~Phillips,
``Matter Effects on Three-Neutrino Oscillations,''
Phys. Rev. D \textbf{22} (1980), 2718
doi:10.1103/PhysRevD.22.2718 

%\cite{Minakata:2021dqh} 
\bibitem{Minakata:2021dqh}
H.~Minakata,
``Symmetry finder: A method for hunting symmetry in neutrino oscillation,''
Phys. Rev. D \textbf{104} (2021) no.7, 075024
doi:10.1103/PhysRevD.104.075024
[arXiv:2106.11472 [hep-ph]]. 

%\cite{Minakata:2021goi} 
\bibitem{Minakata:2021goi}
H.~Minakata,
``Symmetry Finder applied to the 1\textendash{}3 mass eigenstate exchange symmetry,''
Eur. Phys. J. C \textbf{81} (2021) no.11, 1021
doi:10.1140/epjc/s10052-021-09810-5
[arXiv:2107.12086 [hep-ph]]. 

%\cite{Minakata:2022zua}
\bibitem{Minakata:2022zua}
H.~Minakata,
``Symmetry in Neutrino Oscillation in Matter with Non-unitarity,''
Acta Phys. Polon. B \textbf{54} (2023) no.6, 6-A1
doi:10.5506/APhysPolB.54.6-A1
[arXiv:2206.06474 [hep-ph]]. 

%\cite{Minakata:2022yvs} 
\bibitem{Minakata:2022yvs}
H.~Minakata,
``Symmetry in neutrino oscillation in matter: New picture and the $\nu$SM - non-unitarity interplay,'' 
Symmetry \textbf{2022}, 14, 2581
https://doi.org/10.3390/sym14122581
[arXiv:2210.09453 [hep-ph]]. 

%\cite{Minakata:2022pyr}
\bibitem{Minakata:2022pyr}
H.~Minakata,
``Comparative Study of the 1\textendash{}2 Exchange Symmetries in Neutrino Frameworks with Global and Local Validities,''
Acta Phys. Polon. B \textbf{54} (2023) no.4, 3
doi:10.5506/APhysPolB.54.4-A3
[arXiv:2212.06320 [hep-ph]]. 

%\cite{Parke:2018shx}
\bibitem{Parke:2018shx}
S.~Parke,
``Theoretical Aspects of the Quantum Neutrino,''
doi:10.1142/9789811207402\_0008
[arXiv:1801.09643 [hep-ph]].

%\cite{Maki:1962mu}
\bibitem{Maki:1962mu}
Z.~Maki, M.~Nakagawa and S.~Sakata,
``Remarks on the unified model of elementary particles,''
Prog. Theor. Phys. \textbf{28} (1962), 870-880
doi:10.1143/PTP.28.870

%\cite{Minakata:1998bf}
\bibitem{Minakata:1998bf}
H.~Minakata and H.~Nunokawa,
``CP violation versus matter effect in long baseline neutrino oscillation experiments,''
Phys. Rev. D \textbf{57} (1998), 4403-4417
doi:10.1103/PhysRevD.57.4403
[arXiv:hep-ph/9705208 [hep-ph]].

%\cite{Minakata:2015gra} 
\bibitem{Minakata:2015gra}
H.~Minakata and S.~J.~Parke,
``Simple and Compact Expressions for Neutrino Oscillation Probabilities in Matter,''
JHEP \textbf{01} (2016), 180
doi:10.1007/JHEP01(2016)180
[arXiv:1505.01826 [hep-ph]]. 

%\cite{Denton:2016wmg} 
\bibitem{Denton:2016wmg}
P.~B.~Denton, H.~Minakata and S.~J.~Parke,
``Compact Perturbative Expressions For Neutrino Oscillations in Matter,''
JHEP \textbf{06} (2016), 051
doi:10.1007/JHEP06(2016)051
[arXiv:1604.08167 [hep-ph]]. 

%\cite{Martinez-Soler:2019nhb}
\bibitem{Martinez-Soler:2019nhb}
I.~Martinez-Soler and H.~Minakata,
``Perturbing Neutrino Oscillations Around the Solar Resonance,''
PTEP \textbf{2019} (2019) no.7, 073B07
doi:10.1093/ptep/ptz067
[arXiv:1904.07853 [hep-ph]]. 

%\cite{Zaglauer:1988gz}
\bibitem{Zaglauer:1988gz}
H.~W.~Zaglauer and K.~H.~Schwarzer,
``The Mixing Angles in Matter for Three Generations of Neutrinos and the Msw Mechanism,''
Z. Phys. C \textbf{40} (1988), 273
doi:10.1007/BF01555889

%\cite{Minakata:2020oxb}
\bibitem{Minakata:2020oxb}
H.~Minakata,
``Neutrino amplitude decomposition in matter,''
Phys. Rev. D \textbf{103} (2021) no.5, 053004
doi:10.1103/PhysRevD.103.053004
[arXiv:2011.08415 [hep-ph]].

%\cite{Huber:2019frh} 
\bibitem{Huber:2019frh}
P.~Huber, H.~Minakata and R.~Pestes,
``Interference between the Atmospheric and Solar Oscillation Amplitudes,''
Phys. Rev. D \textbf{101} (2020) no.9, 093002
doi:10.1103/PhysRevD.101.093002
[arXiv:1912.02426 [hep-ph]].

%\cite{JUNO:2022mxj}
\bibitem{JUNO:2022mxj}
A.~Abusleme \textit{et al.} [JUNO],
``Sub-percent precision measurement of neutrino oscillation parameters with JUNO,''
Chin. Phys. C \textbf{46} (2022) no.12, 123001
doi:10.1088/1674-1137/ac8bc9
[arXiv:2204.13249 [hep-ex]]. 

%\cite{Minakata:2020ijz}
\bibitem{Minakata:2020ijz}
H.~Minakata,
``Neutrino amplitude decomposition: Toward observing the atmospheric - solar wave interference,''
Eur. Phys. J. C \textbf{80} (2020) no.12, 1207
doi:10.1140/epjc/s10052-020-08746-6
[arXiv:2006.16594 [hep-ph]]. 

%\cite{T2K:2023mcm}
\bibitem{T2K:2023mcm}
K.~Abe \textit{et al.} [T2K],
``Updated T2K measurements of muon neutrino and antineutrino disappearance using 3.6\texttimes{}1021 protons on target,''
Phys. Rev. D \textbf{108} (2023) no.7, 072011
doi:10.1103/PhysRevD.108.072011
[arXiv:2305.09916 [hep-ex]].

%\cite{NOvA:2021nfi}
\bibitem{NOvA:2021nfi}
M.~A.~Acero \textit{et al.} [NOvA],
``Improved measurement of neutrino oscillation parameters by the NOvA experiment,''
Phys. Rev. D \textbf{106} (2022) no.3, 032004
doi:10.1103/PhysRevD.106.032004
[arXiv:2108.08219 [hep-ex]].

%\cite{Hyper-Kamiokande:2018ofw}
\bibitem{Hyper-Kamiokande:2018ofw}
K.~Abe \textit{et al.} [Hyper-Kamiokande],
``Hyper-Kamiokande Design Report,''
[arXiv:1805.04163 [physics.ins-det]].

%\cite{DUNE:2020jqi}
\bibitem{DUNE:2020jqi}
B.~Abi \textit{et al.} [DUNE],
``Long-baseline neutrino oscillation physics potential of the DUNE experiment,''
Eur. Phys. J. C \textbf{80} (2020) no.10, 978
doi:10.1140/epjc/s10052-020-08456-z
[arXiv:2006.16043 [hep-ex]].

%\cite{IceCubeCollaboration:2023wtb} 
\bibitem{IceCubeCollaboration:2023wtb}
R.~Abbasi \textit{et al.} [(IceCube Collaboration)* and IceCube],
``Measurement of atmospheric neutrino mixing with improved IceCube DeepCore calibration and data processing,''
Phys. Rev. D \textbf{108} (2023) no.1, 012014
doi:10.1103/PhysRevD.108.012014
[arXiv:2304.12236 [hep-ex]].

%\cite{KM3NeT:2021ozk} 
\bibitem{KM3NeT:2021ozk}
S.~Aiello \textit{et al.} [KM3NeT],
``Determining the neutrino mass ordering and oscillation parameters with KM3NeT/ORCA,''
Eur. Phys. J. C \textbf{82} (2022) no.1, 26
doi:10.1140/epjc/s10052-021-09893-0
[arXiv:2103.09885 [hep-ex]].

%\cite{Nunokawa:2005nx}
\bibitem{Nunokawa:2005nx}
H.~Nunokawa, S.~J.~Parke and R.~Zukanovich Funchal,
``Another possible way to determine the neutrino mass hierarchy,''
Phys. Rev. D \textbf{72} (2005), 013009
doi:10.1103/PhysRevD.72.013009
[arXiv:hep-ph/0503283 [hep-ph]].

%\cite{Martinez-Soler:2018lcy} 
\bibitem{Martinez-Soler:2018lcy}
I.~Martinez-Soler and H.~Minakata,
``Standard versus Non-Standard CP Phases in Neutrino Oscillation in Matter with Non-Unitarity,''
PTEP \textbf{2020} (2020) no.6, 063B01
doi:10.1093/ptep/ptaa062
[arXiv:1806.10152 [hep-ph]].

%\cite{ParticleDataGroup:2022pth} 
\bibitem{ParticleDataGroup:2022pth}
R.~L.~Workman \textit{et al.} [Particle Data Group],
``Review of Particle Physics,''
PTEP \textbf{2022} (2022), 083C01
doi:10.1093/ptep/ptac097


%\cite{Arafune:1996bt}
\bibitem{Arafune:1996bt}
J.~Arafune and J.~Sato,
``CP and T violation test in neutrino oscillation,''
Phys. Rev. D \textbf{55} (1997), 1653-1658
doi:10.1103/PhysRevD.55.1653
[arXiv:hep-ph/9607437 [hep-ph]].

%\cite{Cervera:2000kp}
\bibitem{Cervera:2000kp}
A.~Cervera, A.~Donini, M.~B.~Gavela, J.~J.~Gomez Cadenas, P.~Hernandez, O.~Mena and S.~Rigolin,
``Golden measurements at a neutrino factory,''
Nucl. Phys. B \textbf{579} (2000), 17-55
[erratum: Nucl. Phys. B \textbf{593} (2001), 731-732]
doi:10.1016/S0550-3213(00)00221-2
[arXiv:hep-ph/0002108 [hep-ph]].

%\cite{Freund:2001pn}
\bibitem{Freund:2001pn}
M.~Freund,
``Analytic approximations for three neutrino oscillation parameters and probabilities in matter,''
Phys. Rev. D \textbf{64} (2001), 053003
doi:10.1103/PhysRevD.64.053003
[arXiv:hep-ph/0103300 [hep-ph]].

%\cite{Akhmedov:2004ny}
\bibitem{Akhmedov:2004ny}
E.~K.~Akhmedov, R.~Johansson, M.~Lindner, T.~Ohlsson and T.~Schwetz,
``Series expansions for three flavor neutrino oscillation probabilities in matter,''
JHEP \textbf{04} (2004), 078
doi:10.1088/1126-6708/2004/04/078
[arXiv:hep-ph/0402175 [hep-ph]]. 

%\cite{Agarwalla:2013tza}
\bibitem{Agarwalla:2013tza}
S.~K.~Agarwalla, Y.~Kao and T.~Takeuchi,
``Analytical approximation of the neutrino oscillation matter effects at large $\theta_{13}$,''
JHEP \textbf{04} (2014), 047
doi:10.1007/JHEP04(2014)047
[arXiv:1302.6773 [hep-ph]]. 

%\cite{Minakata:2021nii}
\bibitem{Minakata:2021nii}
H.~Minakata,
``Toward diagnosing neutrino non-unitarity through CP phase correlations,''
PTEP \textbf{2022} (2022) no.6, 063B03
doi:10.1093/ptep/ptac078
[arXiv:2112.06178 [hep-ph]]. 

%%%%%%%%%%%%% UV %%%%%%%%%%%%%%
%\cite{Antusch:2006vwa} 
\bibitem{Antusch:2006vwa}
S.~Antusch, C.~Biggio, E.~Fernandez-Martinez, M.~B.~Gavela and J.~Lopez-Pavon,
``Unitarity of the Leptonic Mixing Matrix,''
JHEP \textbf{10} (2006), 084
doi:10.1088/1126-6708/2006/10/084
[arXiv:hep-ph/0607020 [hep-ph]].

%\cite{Escrihuela:2015wra}
\bibitem{Escrihuela:2015wra}
F.~J.~Escrihuela, D.~V.~Forero, O.~G.~Miranda, M.~Tortola and J.~W.~F.~Valle,
``On the description of nonunitary neutrino mixing,''
Phys. Rev. D \textbf{92} (2015) no.5, 053009
[erratum: Phys. Rev. D \textbf{93} (2016) no.11, 119905]
doi:10.1103/PhysRevD.92.053009
[arXiv:1503.08879 [hep-ph]].

%\cite{Blennow:2016jkn} 
\bibitem{Blennow:2016jkn}
M.~Blennow, P.~Coloma, E.~Fernandez-Martinez, J.~Hernandez-Garcia and J.~Lopez-Pavon,
``Non-Unitarity, sterile neutrinos, and Non-Standard neutrino Interactions,''
JHEP \textbf{04} (2017), 153
doi:10.1007/JHEP04(2017)153
[arXiv:1609.08637 [hep-ph]]. 

%\cite{Fong:2016yyh} 
\bibitem{Fong:2016yyh}
C.~S.~Fong, H.~Minakata and H.~Nunokawa,
``A framework for testing leptonic unitarity by neutrino oscillation experiments,''
JHEP \textbf{02} (2017), 114
doi:10.1007/JHEP02(2017)114
[arXiv:1609.08623 [hep-ph]].

%\cite{Fong:2017gke}
\bibitem{Fong:2017gke}
C.~S.~Fong, H.~Minakata and H.~Nunokawa,
``Non-unitary evolution of neutrinos in matter and the leptonic unitarity test,''
JHEP \textbf{02} (2019), 015
doi:10.1007/JHEP02(2019)015
[arXiv:1712.02798 [hep-ph]]. 
%%%%%%%%%%%%% UV %%%%%%%%%%%%%%

%\cite{Landau:1981}
\bibitem{Landau:1981}
L.~D.~Landau and E.~M.~Lifshitz, ``Quantum Mechanics: Non-Relativistic Theory'' Third Edition, Butterworth-Heinemann, 1981,
ISBN 978-0750635394

%\cite{Denton:2019ovn}
\bibitem{Denton:2019ovn}
P.~B.~Denton, S.~J.~Parke and X.~Zhang,
``Neutrino oscillations in matter via eigenvalues,''
Phys. Rev. D \textbf{101} (2020) no.9, 093001
doi:10.1103/PhysRevD.101.093001
[arXiv:1907.02534 [hep-ph]].

%\cite{Itzykson:1980rh} 
\bibitem{Itzykson:1980rh}
C.~Itzykson and J.~B.~Zuber, ``Quantum Field Theory,''
McGraw-Hill, 1980,
ISBN 978-0-486-44568-7

%\cite{Parke:2023}
\bibitem{Parke:2023} 
S.~J.~Parke, private communications. 

%\cite{Akhmedov:2009rb}
\bibitem{Akhmedov:2009rb}
E.~K.~Akhmedov and A.~Y.~Smirnov,
``Paradoxes of neutrino oscillations,''
Phys. Atom. Nucl. \textbf{72} (2009), 1363-1381
doi:10.1134/S1063778809080122
[arXiv:0905.1903 [hep-ph]].

%\cite{Akhmedov:2010ua}
\bibitem{Akhmedov:2010ua}
E.~K.~Akhmedov and A.~Y.~Smirnov,
``Neutrino oscillations: Entanglement, energy-momentum conservation and QFT,''
Found. Phys. \textbf{41} (2011), 1279-1306
doi:10.1007/s10701-011-9545-4
[arXiv:1008.2077 [hep-ph]].

\end{thebibliography}
\end{document}